\documentclass[a4paper,11pt]{article}
\pdfoutput=1 

\usepackage{jheppub} 

\usepackage[T1]{fontenc}
\usepackage[mathletters]{ucs}
\usepackage[utf8x]{inputenc}
\usepackage{multicol}
\usepackage{multirow}

\usepackage{booktabs}
\usepackage{colortbl,colordvi,xcolor,color}

\newcolumntype{a}{>{\columncolor{blue}}c}
\PrerenderUnicode{τ}
\PrerenderUnicode{μ}
\PrerenderUnicode{ν}
\PrerenderUnicode{γ}

\title{High Precision Higgs from \\High Energy Muon Colliders}

\preprint{YITP-SB-22-11}
\author{Matthew Forslund}
\author{and Patrick Meade}
\affiliation{C. N. Yang Institute for Theoretical Physics,\\Stony Brook University, Stony Brook, NY 11794}

\emailAdd{matthew.forslund@stonybrook.edu, patrick.meade@stonybrook.edu}

\abstract{Muon colliders are an exciting possibility for reaching the highest energies possible on the shortest timescale.  They potentially combine the greatest strengths of $e^+e^-$ and $pp$ colliders by bridging the energy versus precision dichotomy.  In this paper we study the sensitivity of Higgs properties that can be achieved with a future 3 or 10 TeV muon collider from single Higgs production.  The results presented here represent the first comprehensive picture for the precision achievable including backgrounds and using fast detector simulation with {\sc Delphes}.  Additionally, we compare the results of fast detector simulation with available full simulation studies that include the muon collider specific Beam Induced Background, and show the results are largely unchanged.  We comment on some of the strengths and weaknesses of a high energy muon collider for Higgs physics alone, and demonstrate the complementarity of such a collider with the LHC and $e^+e^-$ Higgs factories.  Furthermore, we discuss some of the exciting avenues for improving future results from both theoretical and detector R\&D that could be undertaken.}

\begin{document} 
\maketitle
% \flushbottom

\section{Introduction}

Since the discovery of the Higgs, there have been two different paths envisioned for future colliders at the energy frontier.  One path focuses on a precision $e^+e^-$ Higgs factory to study its properties with the hope of discovering deviations from beyond the Standard Model (BSM) physics contributions.  The other path imagines pushing the energy frontier as far as possible in the pursuit of the unknown.  This plan was endorsed by the previous US P5~\cite{HEPAPSubcommittee:2014bsm} and recent European Strategy Update~\cite{EuropeanStrategyGroup:2020pow}.  However, to study the Higgs {\em beyond} the important first step of proposed $e^+e^-$ Higgs factories, increasing the center of mass (CM) energy is crucial and muon colliders are a particularly attractive option that we investigate in this paper.

Numerous questions surrounding the Higgs of whether the SM can be fully tested, naturalness, Electroweak Symmetry Breaking (EWSB), flavor,  Higgs portals, or the Electroweak Phase Transition (EWPT), are often deeply intertwined with the number of Higgs particles that can be produced and the signal to background ratio that can be achieved.   For example, in previously proposed $e^+e^-$ Higgs factories $\mathcal{O}(10^6)$ Higgs bosons could be produced~\cite{deBlas:2019rxi}, but there are major SM decay modes of the Higgs with branching fractions of $\mathcal{O}(10^{-8})$ which are simply out of reach of those facilities.  This challenge persists in the context of BSM physics where leading contributions to Higgs observables often decouple as $\mathcal{O}(v^2/m_{BSM}^2)$.  Given the lack of observation of new physics at the LHC thus far, this requires the precision to generically be pushed further in the context of Higgs physics.    Furthermore, if one wants to study the self interactions of the Higgs via multi-Higgs production, sufficient energy and statistics of multi-Higgs production are needed.

Producing more Higgs bosons or multi-Higgs events is controlled by the simple relation that the $N_{ev}\propto \mathcal{L} \sigma$, where $\mathcal{L}$ is the luminosity and $\sigma$ represent the production cross section.  Naively, this implies for Higgs production that the luminosity is the only dial available to increase statistics, given that the mass scale of Higgs physics is known.  However, it has long been known that one can increase the production cross section for Higgs with higher energy colliders, rather than the usual relation that cross sections scale as $1/E_{CM}^2$ above threshold.  For example the LHC and HL-LHC will end up producing about two orders of magnitude more Higgs bosons than the proposed low energy $e^+e^-$ Higgs factories, and a 100 TeV $pp$ collider e.g. FCC-hh could produce $\mathcal{O}(10^{10})$ Higgs particles.  This results from the fact that proton colliders accelerate composite objects, and at high energies the scale of the Higgs moves to lower $x$ which allows one to exploit the growing low-$x$ gluon PDF for Higgs production.  However, this increased production cross section for Higgs at high energies hadron colliders is offset by a poor signal to background ratio.  For hadronic decays of the Higgs in particular, hadron colliders are often ill suited to testing Higgs properties beyond the third generation.  High energy lepton colliders allow for a possible best of both worlds scenario.  

At high energies, the probability of a lepton radiating a forward vector boson increases~\cite{vonWeizsacker:1934nji,Williams:1934ad,Dawson:1984gx}, and in turn the cross section for producing a Higgs via vector boson fusion scales as $\sigma\propto \ln E_{CM}^2$.  When viewed through the lens of generalized parton distribution functions~\cite{Ciafaloni:2005fm,Bauer:2017isx,Han:2020uid,Han:2021kes} this is the analogous phenomena which effectively increases the cross section for Higgs production at high energy hadron colliders.   Additionally, lepton colliders offer high signal to background ratios, especially for hadronic final states.  Nevertheless, logarithmic growth isn't sufficient to parametrically increase statistics unless there is a large jump in energy.  This is well demonstrated from the projections of $e^+e^-$ Higgs factories, where CLIC operating up to 3 TeV doesn't represent a large increase in single Higgs precision measurements compared to low energy options~\cite{deBlas:2019rxi}.    If both the luminosity {\em and} the energy of a lepton collider could be increased, then a high energy lepton collider would be an extremely powerful machine for studying the Higgs and beyond.  Unfortunately, in the case of $e^+e^-$ colliders, increasing the energy and luminosity parametrically are very difficult without technological or financial breakthroughs.  New technologies such as plasma wakefield acceleractors (PWFA) or laser wakefield accelerators (LWFA) are needed to make reaching e.g. the 10 TeV scale practical, which currently are 30+ years away from a practical solution for a collider, as outlined in the European Strategy Accelerator R\&D Roadmap~\cite{Adolphsen:2022bay}.  Furthermore, increasing the luminosity in current state of the art linear collider (LC) designs at the highest energies, such as CLIC, is also very difficult.  For current LC designs $\mathcal{L}\propto$ power consumption, so unless power and financial considerations were to change, CLIC projections are at the limit of what could be accomplished in the next decades . However, high energy muon colliders provide an alternative~\cite{Delahaye:2019omf,Long:2020wfp}.  Muon colliders offer a viable solution to both the energy and luminosity concerns up to at least $\mathcal{O}(10)$ TeV.  Muon collider designs in this energy range roughly have a luminosity which scales as $\mathcal{L}\sim E_{CM}^2$, and possibly even more important $\mathcal{L}/$power consumption {\em increases} with the CM energy, making them a much more efficient option at high energy.  Moreover, designs exist that extend to 10 TeV and have a timeline achievable on the 20 year timescale~\cite{Adolphsen:2022bay}.

Despite the potential advantages of high energy muon colliders from the accelerator side, the physics case has only been rapidly developed over the past few years, as previously there weren't detailed studies exploring $\mathcal{O}(10)$ TeV lepton colliders.  For a broad overview of the physics case see e.g.~\cite{SmashersGuide}.  In this paper, we focus on further developing the sensitivity for single Higgs precision at high energy muon colliders.   Thus far, single Higgs precision studies for high energy muon colliders have been explored only at the signal-only level for most channels~\cite{SmashersGuide} and including backgrounds only for a few specific cases~\cite{Han:2020pif,Bartosik:2020xwr}.  Here we extend the analysis of~\cite{SmashersGuide} to include the hard contributions to backgrounds for all relevant channels using a fast detector simulation.   In particular we focus on two cases, a 10 TeV muon collider with an integrated luminosity of 10/ab and a 3 TeV muon collider with an integrated luminosity of 1/ab.  The 3 TeV option is included as a possible staging option for the International Muon Collider Collaboration (IMCC)~\cite{Aime:2022flm,DeBlas:2022wxr}, and additionally allows us to compare to the currently limited amount of full simulation studies that are available.  This is an important point, as muon colliders present a different type of background from muons which decay in the beam, called Beam Induced Background (BIB).   There are mitigation mechanisms for BIB that we discuss later, but BIB is not included in fast simulation.  Therefore having a point of comparison between fast and full detector simulation is useful to lend credence to the results we present in this paper.

The rest of the paper is structured as follows.  We first review the production cross sections for single Higgs at muon colliders, and the simulation tools used for the hard processes and detector simulation.  We then present results at 3 and 10 TeV for various single Higgs production and decay modes.  To demonstrate the reach of a high energy muon collider we show results for a ``$\kappa$'' fit as a commonly used example.  We then conclude with the general lessons that can be learned from a high energy muon collider for Higgs physics and outline future research directions.  There are other complementary studies for Higgs physics capabilities at future muon colliders.  In particular, multi-Higgs studies looking at the Higgs self-couplings have been explored \cite{Han:2020pif,Chiesa:2020awd} and demonstrate competitive results with 100 TeV $pp$ machines for energies of $\mathcal{O}(10)$ TeV \cite{Mangano:2020sao}.  There is also a potential Higgs program for a muon collider running at the Higgs pole that has been explored in \cite{Barger:1995hr,Barger:1996jm,Han:2012rb}, and most recently in \cite{deBlas:2022aow}.

\section{Higgs production at High Energies and Methodology}\label{sec:methodology}

\begin{figure}[hb!]
    \centering
    \includegraphics[width=\textwidth]{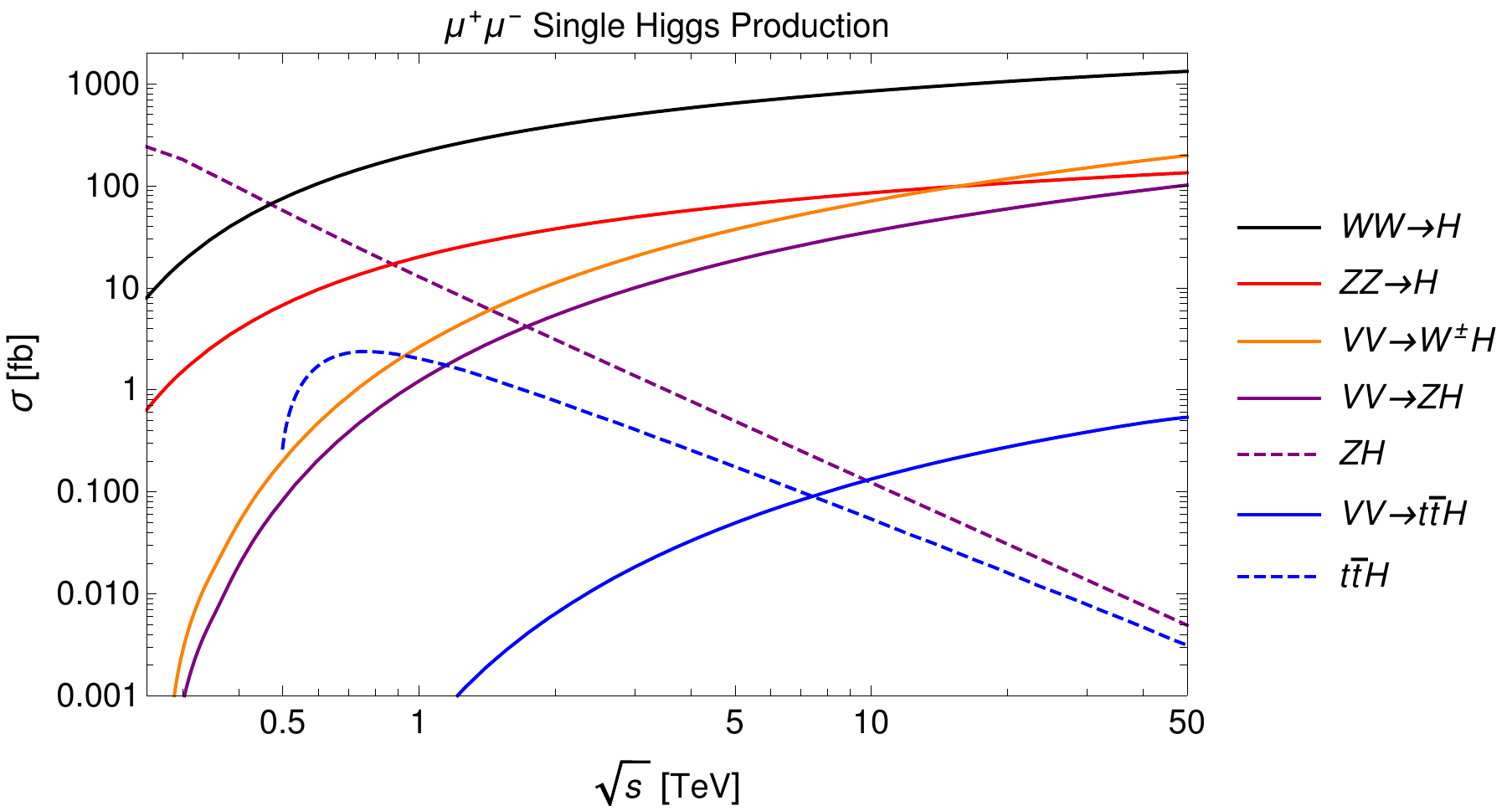}
    \caption{Cross sections for the most important single Higgs production modes as a function of energy. Here $ZH$ and $t\bar{t}H$ are $s$-channel production while the others are vector boson fusion produced in association with any of $(\nu_\mu\bar{\nu}_\mu,\nu_\mu\mu^\pm,\mu^+\mu^-)$.}
    \label{fig:crosssections}
\end{figure}

To understand the basic promise for high energy muon colliders and where particular advances in Higgs precision can be made, it is useful to look at the production cross sections for single Higgs states as a function of CM energy and understand the qualitative lessons that can be learned.  This has been well documented elsewhere, emphasizing the point that at high energy lepton colliders, vector boson fusion (VBF) production modes begin to dominate for many electroweak processes \cite{Buttazzo:2018qqp,Costantini:2020stv}, allowing them to be thought of as gauge boson colliders \cite{Buttazzo:2018qqp,SmashersGuide,Han:2020uid,Han:2021kes}. However, since we focus on single Higgs precision in this paper for muon colliders, in Figure \ref{fig:crosssections} we show only the unpolarized cross sections for the most important production mechanisms as a function of energy, where we have separated $WW$ and $ZZ$ fusion for single Higgs production.   As clearly seen, by 10 TeV, VBF is the dominant production mode for all single Higgs production including $ZH$ and $t\bar{t}H$.  For example, we see that $W^+W^-$ fusion single Higgs overtakes $ZH$ by 500 GeV, $ZZ$ fusion single Higgs overtakes it by 900 GeV, and even VBF $ZH$ production becomes larger by 1.1 TeV.  Since $WW$ provides the largest single Higgs cross section parametrically, it is obvious that high energy muon colliders will provide the most sensitivity to the $hW^+W^-$ coupling compared to other Higgs factory options~\cite{deBlas:2019rxi}.  However, there is also room for complementarity, given that at lower energies $e^+e^-$ colliders rely upon the $hZZ$ coupling for the dominant production mode.  The dominance of VBF and the kinematics of a 10 TeV muon collider presents new challenges as well.  Given that the $ZZ$ and $W^+W^-$ VBF production modes are both large (even though $ZZ$ is clearly subdominant), they serve as a background to each other if there aren't handles to disentangle them.  As we will see in the results of Section~\ref{results}, Higgs being a sizable background for itself is a common feature for high energy muon colliders which benefit from the improved S/B compared to hadron colliders.  

\begin{figure}[h!]
    \centering
    \includegraphics[width=.7\textwidth]{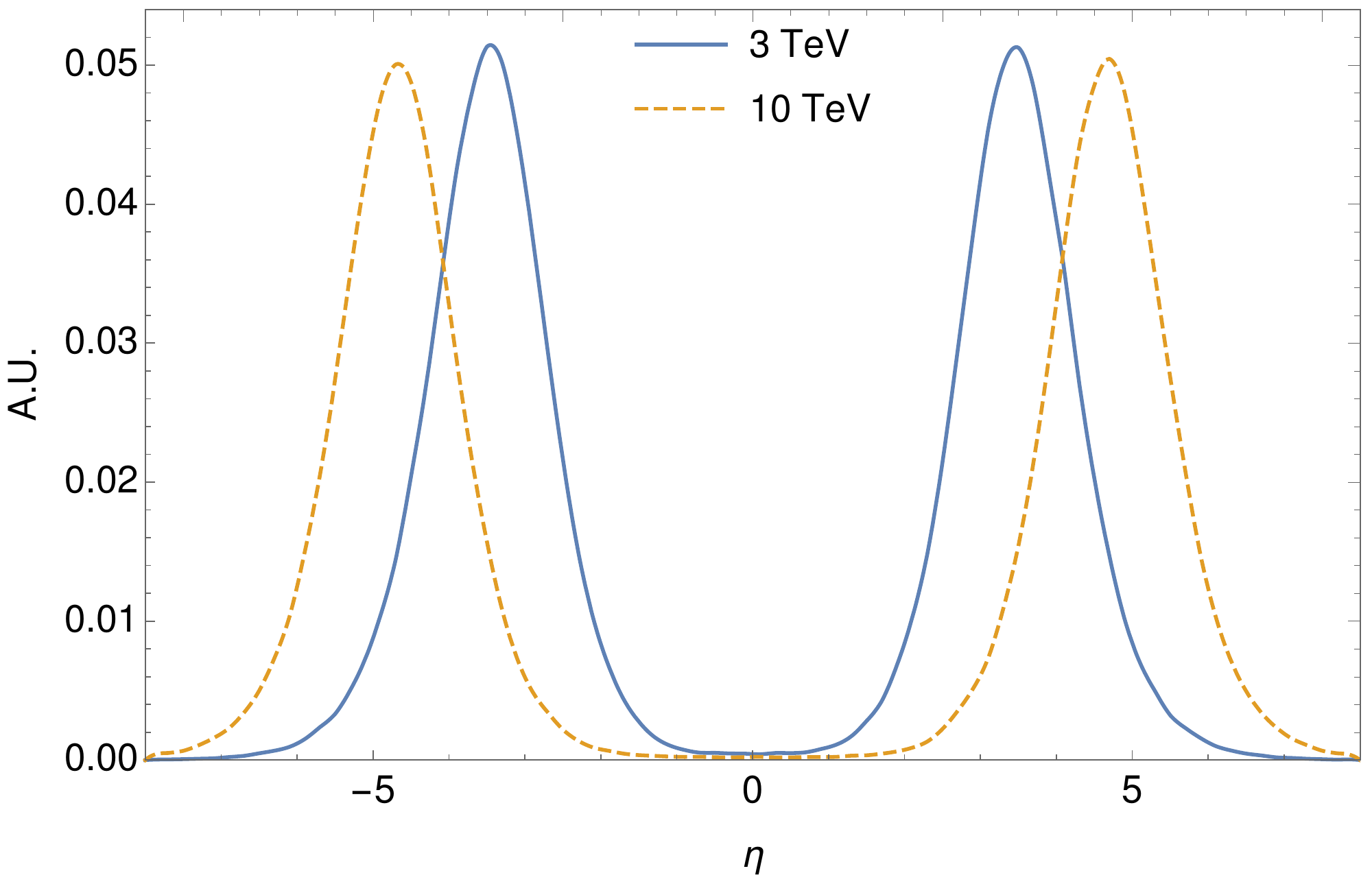}
    \caption{The normalised pseudorapidity distributions of the forward muons from $ZZ$ fusion single Higgs production at 3 and 10 TeV.}
    \label{fig:ForwardMuons}
\end{figure}

An obvious handle to disentangle various VBF production contributions is the ability to tag forward charged particles.  For instance, if one could tag forward muons, one could easily distinguish between $ZZ$ and $W^+W^-$ VBF processes. However, as the $E_{CM}$ increases far beyond $m_h$, the contributions from the charged particles which initiate the VBF production become even more forwards.    In Figure \ref{fig:ForwardMuons}, we show the $\eta$ distribution of forward muons from $ZZ$ fusion production of single Higgs at our two benchmarks of 3 and 10 TeV.  As can be seen for 3 TeV, tagging forward muons isn't particularly challenging compared to the LHC; however, for 10 TeV, it would require a system with capabilities more similar to FCC-$hh$ \cite{FCChhdelphes,FCC:2018vvp} to fully exploit the additional physics gains from tagging forward muons.

\begin{figure}[htb]
    \centering
    \includegraphics[width=.49\textwidth]{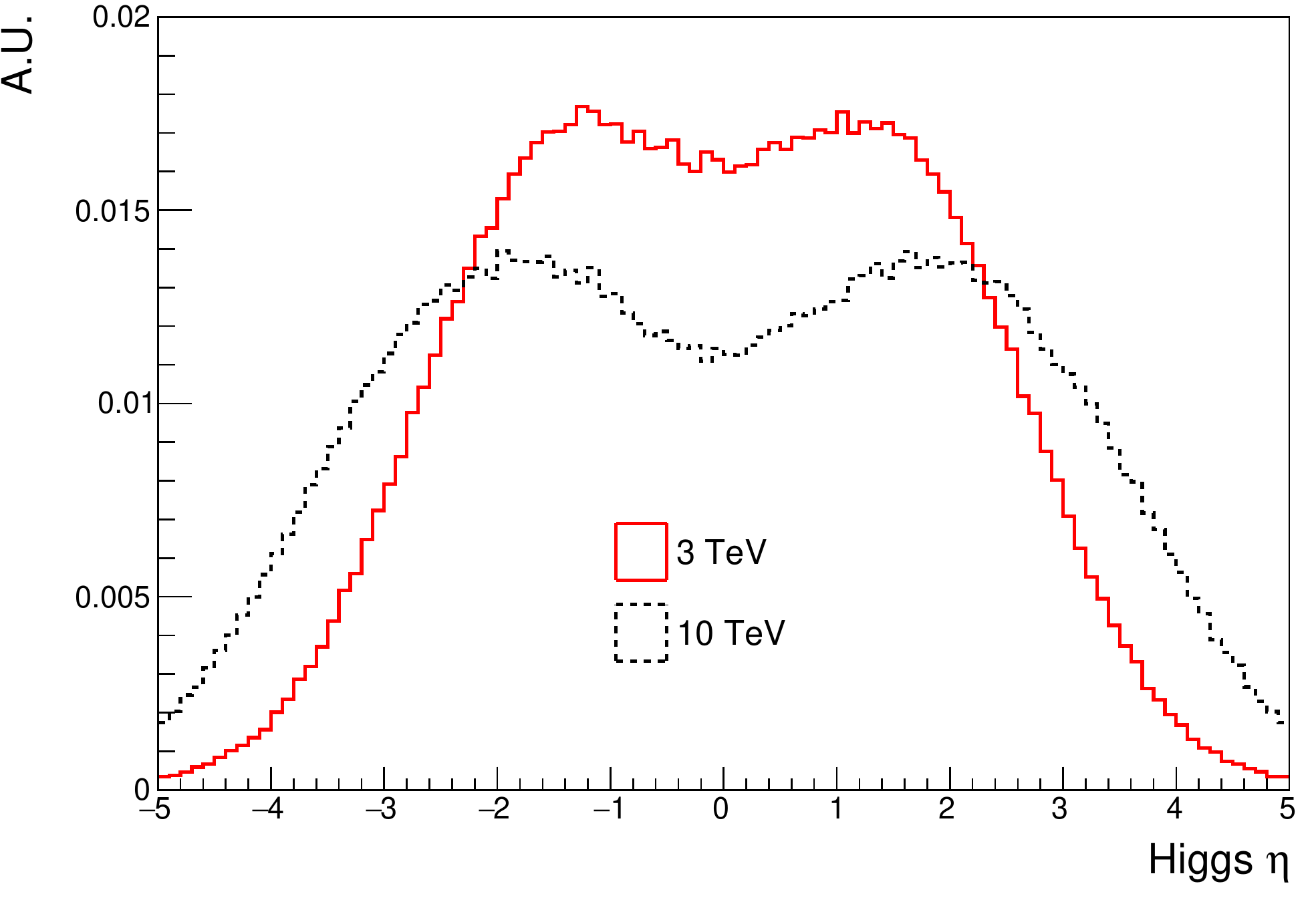}
    \includegraphics[width=.49\textwidth]{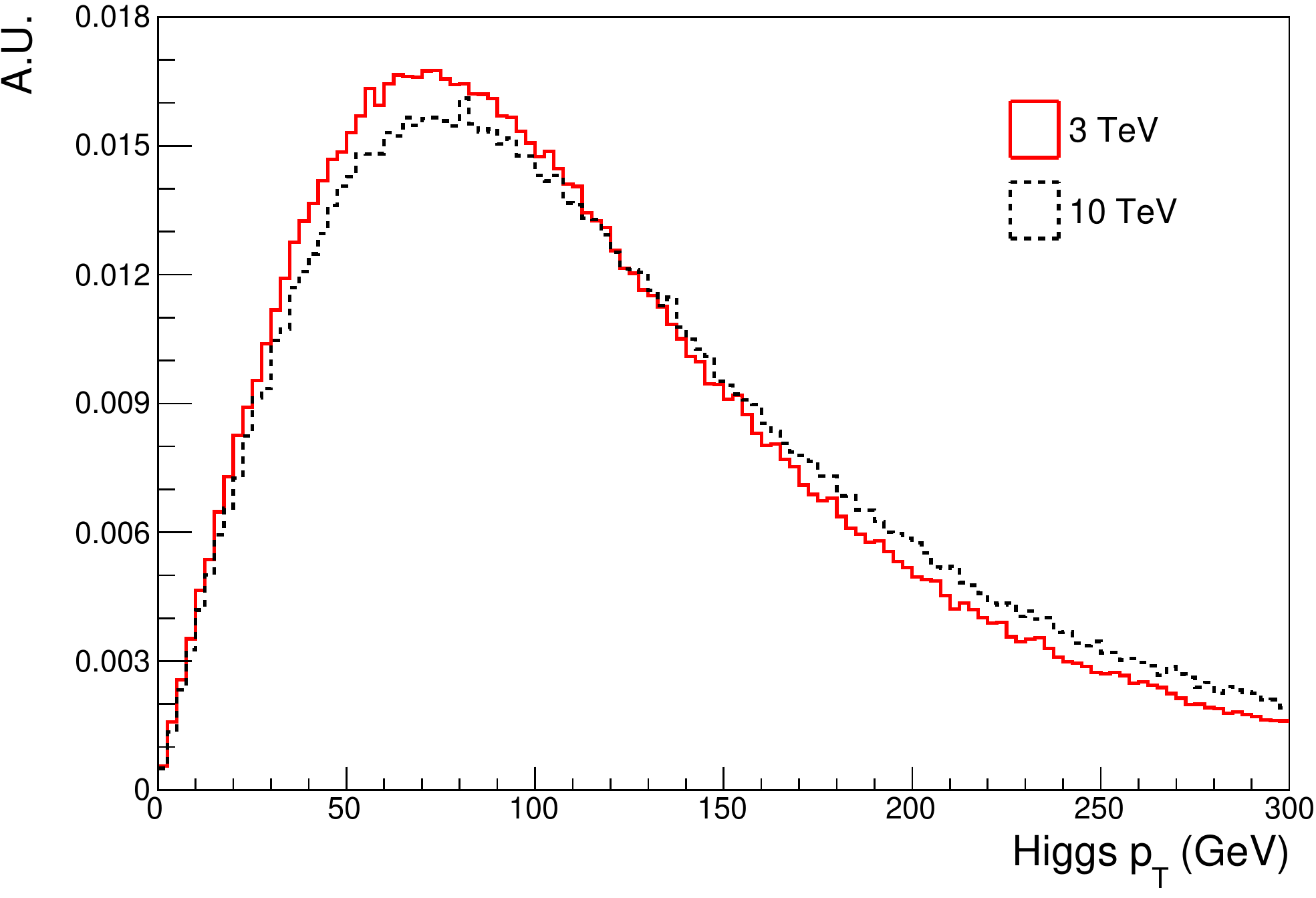}
    \caption{Normalised Higgs pseudorapidity (left) and transverse momentum (right) distributions for $W^+W^-$ fusion events at 3 and 10 TeV.}
    \label{fig:HiggsHists}
\end{figure}

The kinematics of high energy muon colliders are also different for the actual Higgs itself, not just for the VBF byproducts.  As $E_{CM}$ increases, the Higgs becomes more forward, as shown in Figure~\ref{fig:HiggsHists} at 3 and 10 TeV. Compared to the VBF byproducts, the Higgs is not as forward, even at 10 TeV. Nevertheless, depending on the detector design, there can be an interplay between acceptance and Higgs precision.  In particular, the aforementioned BIB has already influenced preliminary detector design ideas.  The exact influence of the BIB depends on the specific accelerator within about 50 meters of the IP; however, mitigation ideas have been proposed for several decades which center around the idea of introducing Tungsten nozzles near the interaction point~\cite{Foster:1995ru,Bartosik:2019dzq,Mokhov:2011zzb}.  Of course, these nozzles reduce the ability to extend a single detector to the very forward region to maintain ``$4\pi$'' coverage, and as an example, for nozzles optimized for $E_{CM}=1.5$~TeV, this would impede coverage more forward than $\eta \sim 2.5$ \cite{Mokhov:2011zzb,Mokhov:2011zzd}. However, the BIB mitigation optimization depends on the CM energy and accelerator design so therefore the tradeoff with acceptance for physics targets needs further study. Moreover, there can be more modern techniques used in conjunction such as precision timing detectors like at the HL-LHC \cite{CMS_MTD,ATLAS_HGTD} that should be able to further mitigate the effects of the BIB beyond what has been studied thus far. A summary of the current status of muon detector design and full simulation can be found in~\cite{Bartosik:2022ctn,Jindariani:2022gxj}. We do not include any effects BIB in our simulation assuming they can be sufficiently mitigated.  However, in Section~\ref{results}, we will show what the effects are on a study that has included BIB in full simulation to calibrate our fast simulation studies which neglect it.  Transverse kinematics of the Higgs are relatively unaffected by increasing $E_{CM}$, as can be seen in Figure~\ref{fig:HiggsHists}  for the $p_T$ distribution of the Higgs at 3 and 10 TeV.  However, as differential observables are explored in more detail in future work, these small differences should be explored further.

\subsection{MC simulation of Higgs and Backgrounds}

As stated earlier, we consider two of the future muon collider energy/luminosity benchmarks set by the Snowmass muon collider forum \cite{MuCollForum}: 3 TeV with total integrated luminosity of 1/ab, and 10 TeV with total integrated luminosity of 10/ab, which also coincides with current IMCC plans. Event generation is done at leading order using {\sc MadGraph5} \cite{Alwall:2014} with parton showering handled by {\sc Pythia8} \cite{Sjostrand:2014zea}. For most channels, the dominant backgrounds are $2\rightarrow4$ processes of the form $\mu^+\mu^-\rightarrow (\nu_\mu\bar{\nu}_\mu,\nu_\mu\mu^\pm,\mu^+\mu^-)f\bar{f}$ where $f$ is a channel-specific fermion. Just like for Higgs production, these processes are dominated by VBF at high energies and therefore the $\nu_\mu,\mu^\pm$ in the final state are usually very forward, resulting in very similar kinematics to our signal. We also include additional contributions from diboson production such as $\mu^+\mu^-\rightarrow (\nu_\mu\bar{\nu}_\mu,\nu_\mu\mu^\pm,\mu^+\mu^-)VV \rightarrow (\nu_\mu\bar{\nu}_\mu,\nu_\mu\mu^\pm,\mu^+\mu^-)f\bar{f}f\bar{f}$ which contribute the vast majority of the full $2\rightarrow 6$ cross section for the same final state. However, for $H\rightarrow WW^*,ZZ^*$ final states, we generate the full $2\rightarrow6$ processes instead since the off-shell regions matter more. In addition to process-dependent generation cuts that are kept softer than analysis cuts, all background processes with muons in the final state are generated with $p_{T_\mu}>10$ GeV and $\Delta R_{\ell\ell}>0.01$ to avoid phase space singularities from $W\gamma/Z\gamma/\gamma\gamma$-fusion. The more complicated $2\rightarrow 6$ processes are generated with $p_{T_{\ell,j}}>10$ GeV, $10<m_{jj}<300$ GeV, and $\Delta R_{\ell\ell,\ell j, jj}>0.1$. 

Throughout this work we use branching ratios from the CERN Yellow Report \cite{LHCHiggsCrossSectionWorkingGroup:2013rie}. We generally use the default parameters in {\sc MadGraph5}, with the Higgs width set to $\Gamma_H=4.07$ MeV and $m_H=125$ GeV. The precision for each channel in section \ref{results} is estimated simply using $\frac{\Delta\sigma}{\sigma} = \frac{\sqrt{S+B}}{S}$, where $S$ and $B$ are the number of signal events and background events satisfying the analysis cuts. The most important backgrounds for each channel are generally similar to CLIC \cite{Abramowicz:2016zbo}, not surprisingly since it is the closest $e^+e^-$ collider to our setup. However, we do not have the bremsstrahlung induced backgrounds that contribute significantly at high energy $e^+e^-$ colliders.  Additional ``collinear'' backgrounds from $q/g$ components of the muon~\cite{Han:2021kes} are not considered here, although we comment further on them in Section~\ref{sec:conclusions}. Likewise, we do not include collinear backgrounds from low-virtuality $\gamma$'s, which are potentially very important. We leave the impact of these collinear $\gamma/q/g$ induced backgrounds to future work. 

We additionally do not include the effects of initial state radiation (ISR), which should be included in fully comprehensive future studies. We note, however, that it has been checked for some channels that including ISR with {\sc Whizard} \cite{whizard,omega} while employing proper cuts for processes with final state photons~\cite{Kalinowski:2020lhp} yields similar results.

\subsection{Detector Simulation}\label{sec:fastsim}

We use {\sc Delphes} \cite{deFavereau:2013fsa} fast simulation to model the detector reconstruction and performance. The card used is the muon collider detector card \cite{delphesTalk} included with the latest {\sc Delphes} releases, which is mostly a hybrid of the CLIC \cite{CLICdelphes} and FCC-$hh$ \cite{FCChhdelphes} cards and is the same card as used in the earlier signal-only study in \cite{SmashersGuide}. We use this card at both of the studied energy/luminosity benchmarks. The card has a jet $p_T$ resolution of 2\% for $|\eta|<0.76$ and 5\% for $|\eta|>0.76$. In addition, a cutoff of $|\eta|<2.5$ for all detected particles is added to approximate the previously discussed tungsten nozzles with an opening angle of $\approx 10^\circ$ at 1.5 TeV. This nozzle opening angle should be able to be reduced at higher energies since the relevant radiation is more forward \cite{Chiesa:2020awd}, but since detailed BIB studies above 1.5 TeV have not yet been done, we use this default as a conservative starting point.

Bottom quark tagging is done using CLIC's tight working point \cite{CLICdelphes}, which consists of a flat $50$\% $b$-tagging efficiency with energy and $\eta$-dependent mistagging rates ranging from 0.07\%-3\% for $c$-quarks and 0.02\%-0.6\% for light quarks, respectively. This corresponds well to the existing conservative full-simulation $b\bar{b}$ studies \cite{Bartosik:2020xwr,Bartosik:2022ctn}, which can serve as a potential floor for $b$-tagging. Since the {\sc Delphes} card does not include $c$-tagging by default, we use flat rates inspired by ILC \cite{Suehara:2016,ILDConceptGroup:2020sfq}. We choose a 20\% working point, with flat mistagging rates of 1.3\% for $b$-jets and 0.66\% for light jets corresponding to the curves in Figure 2 of \cite{Suehara:2016}. These tagging rates clearly do not serve as a final say, but rather just as a reasonable starting point.

An additional hypothetical forward muon detector in the region $|\eta|>2.5$ with 90\% ($0.5<p_T<1$ GeV) - 95\% ($p_T>1$ GeV) efficiency is included in the {\sc Delphes} card. Such a detector would be important for distinguishing between $W^+W^-$ fusion and $ZZ$ fusion processes, as discussed already. However, any specific forward muon detector would necessitate a detector design, which does not yet exist for the energies we study. We therefore take inspiration from the proposed FCC-$hh$ \cite{FCChhdelphes,FCC:2018vvp} and consider forward muon tagging up to $|\eta|\leq6$, without assuming anything about the momentum resolution of such a detector. Additional results for most channels are shown using this detector by incorporating a $N_\mu = 0$ cut for $W^+W^-$ fusion and a $N_\mu = 2$ cut for $ZZ$ fusion.

Jet clustering is performed using the Valencia (VLC) jet clustering algorithm \cite{Boronat:2014hva,Boronat:2016tgd} with $\beta=\gamma=1$, usually used in exclusive mode. This algorithm is a generalization of longitudinally invariant $e^+e^-$ clustering algorithms with separate beam and particle distance parameters based on energy and angle. The algorithm has been found to perform well at high energy lepton colliders where $\gamma\gamma\rightarrow hadrons$ becomes an important background, especially in the forward regions.

%%%%%%%%%%%%%%%%%%
%%%%%%%%%%%%%%%%%%
\section{Results}\label{results}

In this section we compile results for all major decay modes of the Higgs. We first consider the combined VBF production mode for each channel. Within each subsection we also consider the ability to separate production modes of VBF. This lets us demonstrate the effects of having the ability to tag forward muons and separate $W^+W^-$ fusion from $ZZ$ fusion, greatly improving the sensitivity to the $hZZ$ coupling using muon colliders alone. Additionally, we include a separate subsection for $t\bar{t}H$ production to show the sensitivity of high energy muon colliders to $y_t$.   This is particularly novel in the case of a 10 TeV muon collider, where VBF production of $t\bar{t}H$  would dominate, which is different than previous studies at $e^+e^-$ colliders with lower energy.  All studies in this section are performed using the MC and fast simulation as outlined in Section~\ref{sec:methodology}.  However, in the case of $b\bar{b}$ production at 3 TeV, there is a corresponding full simulation study including BIB~\cite{Bartosik:2022ctn} that we show the comparison to in Section~\ref{fullsim}.

\subsection{\texorpdfstring{$b\bar{b}$, $c\bar{c}$, $gg$}{bb, cc, gg}}\label{wwf-jj}

We begin by looking at the two-body hadronic Higgs decays $b\bar{b},c\bar{c},$ and $gg$, considering each channel independently. Events are clustered in exclusive mode into two VLC jets with $R=0.5$.  For the $H\rightarrow b\bar{b}$ channel, prior to analysis cuts, we first apply an additional correction to $b$-tagged jets at the analysis level beyond what is already done in the {\sc Delphes} card. This correction is the same one used in~\cite{Homiller:2019} and is a rough approximation to the correction used by ATLAS~\cite{ATLAS:2017}. The correction smoothly scales the 4-momentum of the $b$-tagged jets by up to $\sim 1.08$ at low $p_T$ to account for energy losses due to neutrinos, and yields a $m_H$ peak centered near 125 GeV.

\begin{figure}[h]
    \centering
    \includegraphics[width=.49\textwidth]{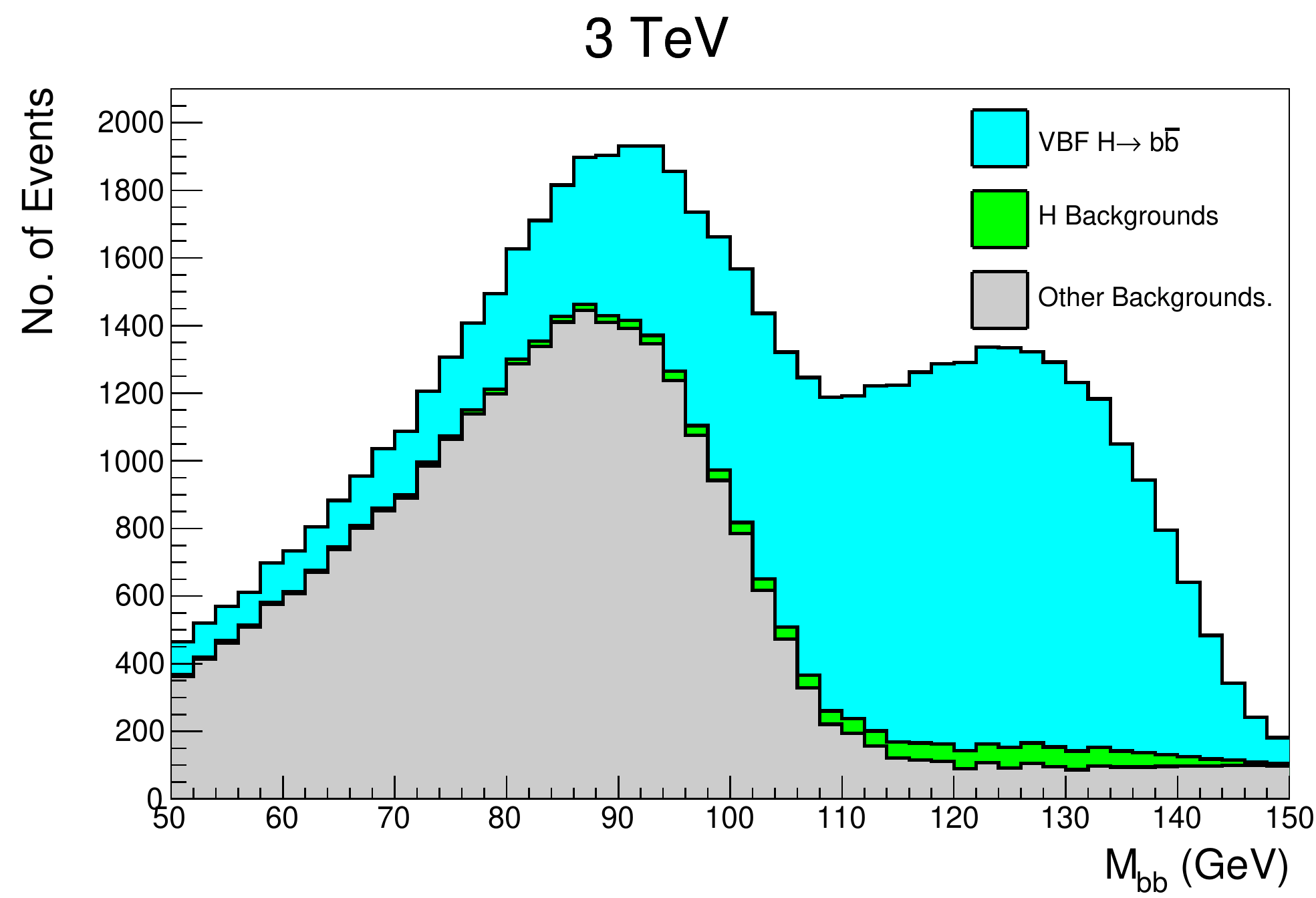}
    \includegraphics[width=.49\textwidth]{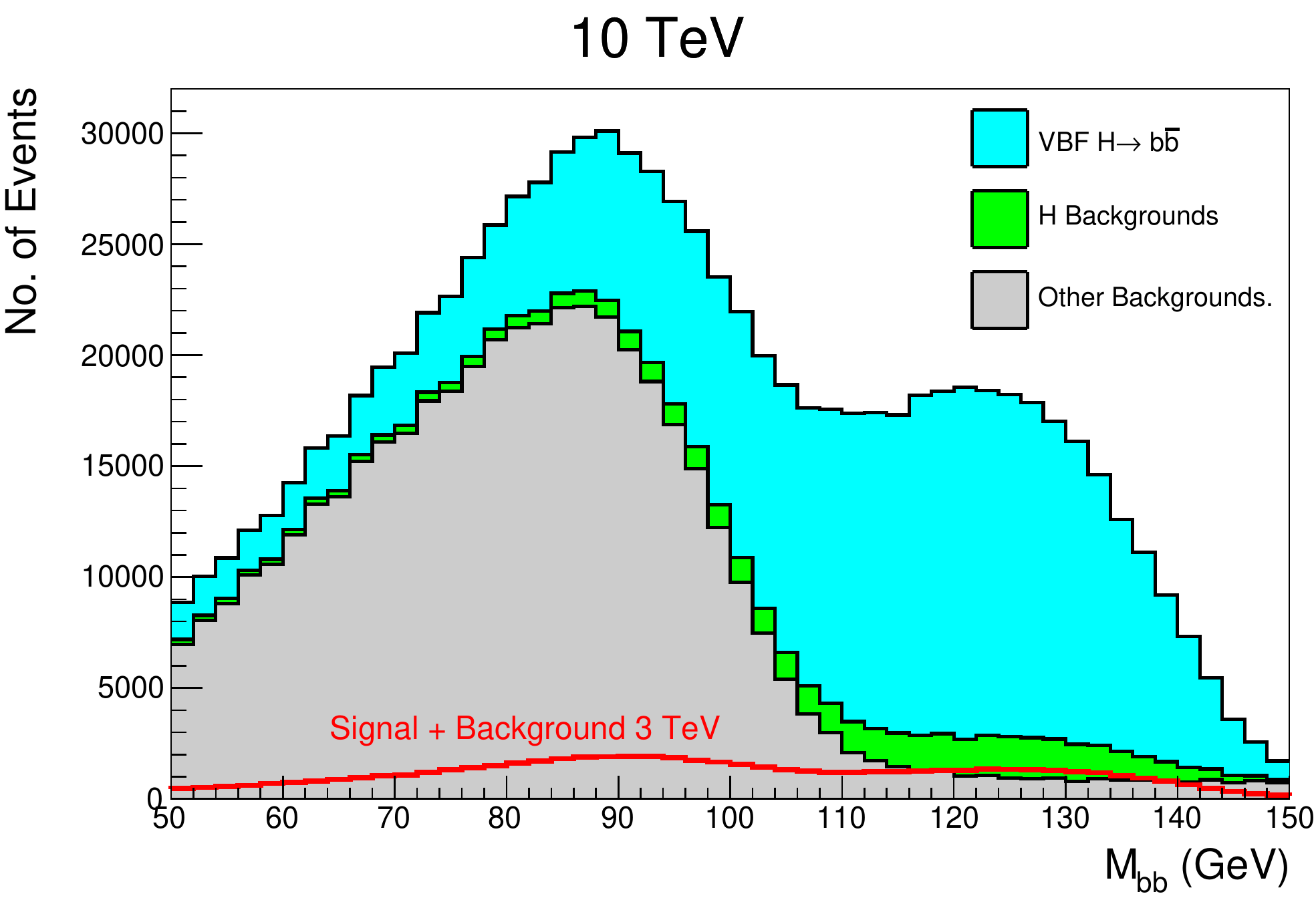}
    \caption{Stacked $b\bar{b}$ pair invariant mass histograms for the $H\rightarrow b\bar{b}$ analysis after $p_T$ cuts and $b$-tagging at (left) 3 TeV and (right) 10 TeV. The sum of signal and background at 3 TeV is overlayed on the 10 TeV plot for ease of comparison.}
    \label{fig:bbHists}
\end{figure}

After applying this correction and $b$-tagging, we select events with two $b$-tagged jets with $p_T>40$ GeV. The invariant mass of the two jets is required to be in the range $100<m_H<150$ GeV. Since this channel has such a large branching ratio, there are already many more signal events than background after applying this cut, and so we do not find it necessary to apply any additional cuts. The dominant background is the upper tail of the $Z$ resonance in $\mu^+\mu^-\rightarrow\nu_\mu\bar{\nu}_\mu j j$, contributing over 60\% of the background events alone. We show the stacked histograms of signal and backgrounds at 3 and 10 TeV after the $p_T$ cut and $b$-tagging in Figure~\ref{fig:bbHists}. We find a precision of 0.76\% at 3 TeV and 0.21\% at 10 TeV for the combined VBF production mode.

For distinguishing the $W^+W^-$ and $ZZ$ fusion production modes, we apply the same set of cuts, seperating between the modes using forward muon tagging up to $|\eta|\leq6$ using the detector discussed in Section~\ref{sec:fastsim}. We require at least two forward muons for $ZZ$ fusion and no detected muons for $W^+W^-$ fusion. The largest background for $W^+W^-$ fusion remains the same, while the largest background for $ZZ$ fusion becomes $(\nu_\mu\mu^\pm)W^\mp H$, especially at 10 TeV where it contributes over half of the background events alone. The final precision we find with this cut for $W^+W^-$ fusion is 0.80(2.6)\% at 3 TeV and 0.22(0.77)\% at 10 TeV for the $W^+W^-$($ZZ$) fusion production mode.

\begin{table}[t]
    \centering
    \renewcommand{\arraystretch}{1.3}
    \setlength{\arrayrulewidth}{.3mm}
    \setlength{\tabcolsep}{0.5 em}
    \begin{tabular}{|c||c|c|c||c|c|c|}
    \multicolumn{1}{c}{} & \multicolumn{6}{c}{Number of Events} \\
    \hline
    \multirow{2}{*}{Process} & 
    \multicolumn{3}{c}{$3\,\textrm{TeV}$} & \multicolumn{3}{c|}{$10\,\textrm{TeV}$}\\ \cline{2-7}
    & $b\bar{b}$ & $c\bar{c}$ & $gg$ & $b\bar{b}$ & $c\bar{c}$ & $gg$ \\ \hline \hline    
        $\mu^+\mu^-\rightarrow \nu_\mu \bar{\nu}_\mu H; \ H\rightarrow X$ & 19000 & 154 & 8570 & 251000 & 2030 & 125000 \\
        $\mu^+\mu^-\rightarrow \mu^+\mu^- H; \ H\rightarrow X$ & 2000 & 16 & 951 & 26700 & 220 & 14300 \\
        \hline
        $\mu^+\mu^-\rightarrow (\mu^+\mu^-,\nu_\mu\bar{\nu}_\mu) H; \ H\not\rightarrow X$ & 75 & 52 & 23400 & 1310 & 1040 & 339000\\
        $\mu^+\mu^-\rightarrow (\mu^+\mu^-,\nu_\mu\bar{\nu}_\mu) jj$  & 2760 & 183 & 24900 & 34700 & 2300 & 355000\\
        $\mu^+\mu^-\rightarrow \nu_\mu\mu^\pm jj$  & 3 & 20 & 18200 & 93 & 718 & 283000 \\
        Others  & 1440 & 70 & 21800 & 37000 & 1610 & 412000 \\
        \hline
        Total Backgrounds  & 4280 & 325 & 88300 & 73200 & 5670 & 1390000 \\
        \hline
    \end{tabular}
    \caption{Signal and some of the most important backgrounds for VBF $H\rightarrow X$, with $X$ one of $(b\bar{b},c\bar{c},gg)$, after applying flavor tagging and analysis cuts. ``Others'' consists of $s$-channel and VBF diboson production, $tb$, and $t\bar{t}$. }
    \label{tab:hadronicbackgrounds}
\end{table}

A very similar analysis is done for $c\bar{c}$. A modified, weaker version of the $p_T$ correction discussed for $b\bar{b}$ with the same $p_T$ dependence is applied to $c$-jets. We require two $c$-tagged jets, both with corrected $p_T>40$ GeV. The invariant mass cut we apply is $105<m_H<145$ GeV, tighter than for $b\bar{b}$ due to the smaller branching ratio for $c\bar{c}$ and slightly different invariant mass distribution. In addition to the analogous backgrounds to those in $b\bar{b}$, $H\rightarrow b\bar{b}$ mistagged as $c\bar{c}$ becomes the second largest background for this process. We obtain a precision of 13\% at 3 TeV and 4.0\% at 10 TeV after applying these cuts for the combined VBF production modes. When requiring either $N_\mu = 0$ or $N_\mu = 2$ with forward muon tagging, we find a precision of 12(72)\% at 3 TeV and 3.6(17)\% at 10 TeV for $W^+W^-$($ZZ$) fusion, respectively.

For $gg$, despite having a substantially larger branching ratio than $c\bar{c}$, the lack of signal tagging information makes the backgrounds substantially larger. We select events with two jets with $p_T>40$ GeV that were not tagged as $c$-jets or $b$-jets. We do not apply any $p_T$ correction to these jets, and as a result the Higgs peak is slightly shifted below 125 GeV. We then apply a two-jet invariant mass cut of $95<m_H<135$ GeV as the only analysis cut. For this channel, backgrounds from other hadronic Higgs decays contribute the most, with similarly large contributions from $Z$ decays in $\mu^+\mu^-\rightarrow\nu_\mu\bar{\nu}_\mu jj$ and contributions from $W$ decays in $\mu^+\mu^-\rightarrow\mu^\pm\nu_\mu jj$, which were mostly removed via tagging information in the other hadronic channels. We find a precision of 3.28\% at 3 TeV and 0.89\% at 10 TeV for the combined VBF signal. Incorporating forward muon tagging information as before, we obtain 2.8(14)\% at 3 TeV and 0.79(3.3)\% at 10 TeV for $W^+W^-$($ZZ$) fusion.

A summary of the signal and most important backgrounds without forward tagging for $b\bar{b}$, $c\bar{c}$, and $gg$ is shown in Table~\ref{tab:hadronicbackgrounds}.

\subsubsection{Comparison to full sim and BIB}\label{fullsim}

\begin{figure}[hbt]
    \centering
    \includegraphics[width=.6\textwidth]{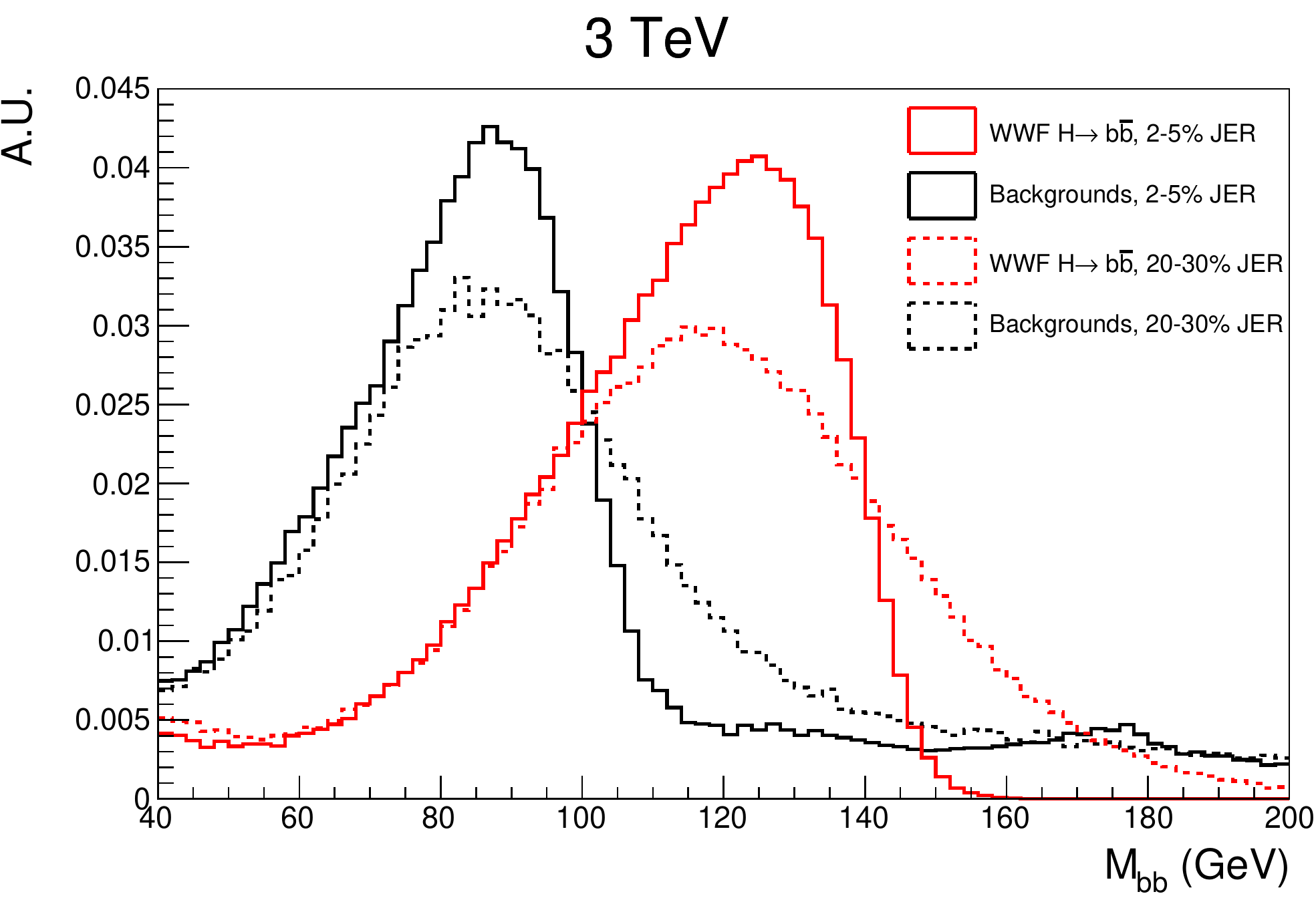}
    \caption{Normalised invariant mass distributions of $b\bar{b}$ pairs for the combined VBF signal (red) and the combined backgrounds (black) with the {\sc Delphes} card default 2-5\% jet energy resolution (solid) and reduced 20-30\% (dashed) jet energy resolution coming from current full simulation in the presence of BIB \cite{Bartosik:2022ctn}.}
    \label{fig:fullsimJER}
\end{figure}

Ideally all the studies presented in this paper would be verified with full simulation including BIB.  However, for a muon collider to even simulate the BIB requires an understanding of the accelerator design and the machine detector interface (MDI) which is much more correlated than for $e^+e^-$ or $pp$ colliders.  Currently, the state of the art simulation from the IMCC includes BIB simulated at $E_{CM}=1.5$~TeV; however, physics studies with full simulation at 3 TeV are then overlaid with the 1.5 TeV BIB~\cite{Bartosik:2022ctn}.  Given that the BIB contribution should become more forward at higher CM energy, this allows for a conservative estimate of the effects of BIB at 3 TeV.  While we cannot directly overlay BIB with our fast simulation results, we can understand how the basic properties that enter our fast simulation compare to the current detector performance shown in~\cite{Bartosik:2022ctn}.  One of the largest difference that arises between our fast simulation and the effects of BIB in full simulation for hadronic events is the Jet Energy Resolution (JER).  Compared to the resolution discussed in Section~\ref{sec:fastsim}, full simulation is almost an order of magnitude worse and ranges from 20-30\% for most of the $p_T$ range of interest~\cite{Bartosik:2022ctn}.  For the $H\rightarrow b\bar{b}$ study, we have modified {\sc Delphes} to account for this worse JER. The resultant normalised signal and total background invariant mass histograms are shown in Figure \ref{fig:fullsimJER}, overlayed with the same results using the {\sc Delphes} default of 2-5\% for comparison. We change the invariant mass cut to $100<m_H<200$ GeV to account for the additional spread, yielding a precision for total VBF at 3 TeV of 0.86\% compared to 0.76\% with the {\sc Delphes} default for the muon collider.  The rudimentary b-tagging implemented in~\cite{Bartosik:2022ctn} has an efficiency very similar to the {\sc Delphes} default, however the light jet mistag rate is also worse for full simulation at this point.  Nevertheless, given the backgrounds shown in Table~\ref{tab:hadronicbackgrounds} and the mistag rate given in~\cite{Bartosik:2022ctn}, this will give smaller effects to the precision than the JER change. Clearly, even in this extremely unoptimized version of the detector, MDI, and analysis there is very little effect on the physics and there is a great deal of room for improvement.  With this data point in mind, it lets us at least calibrate that the results of our fast simulation studies and validate that using our fast simulation defaults are not overly speculative.

\subsection{\texorpdfstring{$\tau^+\tau^-$}{ττ}}\label{wwf-tautau}

\begin{table}[t]
    \centering
    \renewcommand{\arraystretch}{1.3}
    \setlength{\arrayrulewidth}{.3mm}
    \setlength{\tabcolsep}{0.6 em}
    \begin{tabular}{|c||c|c|c||c|c|c|}
    \hline
    \multirow{2}{*}{Process} & 
    \multicolumn{3}{c}{$3\,\textrm{TeV}$} & \multicolumn{3}{c|}{$10\,\textrm{TeV}$}\\ \cline{2-7}
    & $\sigma \ (\textrm{fb})$ & $\epsilon \ (\%)$ & N & $\sigma \ (\textrm{fb})$ & $\epsilon\  (\%)$ & N \\ \hline \hline    
        $\mu^+\mu^-\rightarrow \nu_\mu \bar{\nu}_\mu H; \ H\rightarrow\tau^+\tau^-$ & 31.5 & 4.0 & 1240 & 53.1 & 3.3 & 17500 \\
        $\mu^+\mu^-\rightarrow \mu^+ \mu^- H; \ H\rightarrow\tau^+\tau^-$ & $3.2$ & $4.3$ & 139 & 5.4 & 3.5 & 1910 \\
        \hline
        $\mu^+\mu^-\rightarrow \nu_\mu \bar{\nu}_\mu \tau^+\tau^-$ & 274 & 0.38 & 1050 & 426 & 0.33 & 14000\\
        $\mu^+\mu^-\rightarrow (\mu^+\mu^-,\nu_\mu\bar{\nu}_\mu) H; \ H\not\rightarrow \tau^+\tau^-$ & 514 & 0.052 & 265 & 867 & 0.055 & 4810 \\
        Others & - & - & 384 & - & - & 9100\\
        \hline
    \end{tabular}
    \caption{Signal and most important backgrounds for combined VBF $H\rightarrow \tau^+\tau^-$ after applying $\tau$-tagging and analysis cuts. Here ``Others'' consists of $s$-channel and VBF diboson production, $tb$, and $t\bar{t}$.}
    \label{tab:tautaubackgrounds}
\end{table}

For $H\rightarrow \tau^+\tau^-$, the branching ratio of $6.32\%$ yields a cross section of 31.5 fb at 3 TeV and 53.1 at 10 TeV, making it one of the largest channels. Events are clustered in exclusive mode into two jets using the VLC algorithm with $R=0.5$. $\tau$-tagging is done using the default working point in the {\sc Delphes} card, which has an $80$\% efficiency for hadronic $\tau$-decays with a $2$\% jet mistagging rate and $0.1$\% electron mistagging rate. 

Events with two $\tau$-tags are then required to have two jets satisfying $|\eta| < 2.5$, $p_T > 40$ GeV with a two jet invariant mass $80<m_{\tau\tau}<130$ GeV. An additional cut is applied on the angle between the two jets of $\theta_{\tau\tau}>15$ ($3\,\textrm{TeV}$) or $\theta_{\tau\tau}>20$ ($10\,\textrm{TeV}$) to reduce the $\mu^+\mu^-\rightarrow(\mu^+\mu^-,\nu_\mu\bar{\nu}_\mu)\tau^+\tau^-$ background coming mostly from $Z$ decays. Since all VBF fusion events inherently carry a large amount of missing energy either from the neutrinos in $W^+W^-$ fusion or forward muons disappearing down the beampipe in $ZZ$ fusion, the missing energy from $\tau$ decays is not found to be helpful in distinguishing signal from background. Even after applying these cuts, $\mu^+\mu^-\rightarrow(\mu^+\mu^-,\nu_\mu\bar{\nu}_\mu)\tau^+\tau^-$ remains the dominant background, contributing over half of the background events alone. A summary of the signal and most relevant backgrounds can be found in Table \ref{tab:tautaubackgrounds}, yielding a precision of $4.0$\% at 3 TeV and $1.1$\% at 10 TeV. When incorporating forward tagging capabilities to differentiate the production modes, the resultant precision is 3.8(21)\% at 3 TeV and 1.1(4.8)\% at 10 TeV for $W^+W^-$($ZZ$) fusion.

\subsection{\texorpdfstring{$WW^*$}{WW*}}\label{wwf-ww}

\begin{table}[t]
    \centering
    \renewcommand{\arraystretch}{1.3}
    \setlength{\arrayrulewidth}{.3mm}
    \setlength{\tabcolsep}{0.53 em}
    \begin{tabular}{|c||c|c|c||c|c|c|}
    \hline
    \multirow{2}{*}{Process} & 
    \multicolumn{3}{c}{$3\,\textrm{TeV}$} & \multicolumn{3}{c|}{$10\,\textrm{TeV}$}\\ \cline{2-7}
    & $\sigma \ (\textrm{fb})$ & $\epsilon \ (\%)$ & N & $\sigma \ (\textrm{fb})$ & $\epsilon\  (\%)$ & N \\ \hline \hline    
        $\mu^+\mu^-\rightarrow \nu_\mu \bar{\nu}_\mu H; \ H\rightarrow WW^*\rightarrow 4j $& 52.3 & 2.22 & 1160 & 88.2 & 3.03 & 26700 \\
        $\mu^+\mu^-\rightarrow \mu^+ \mu^- H; \ H\rightarrow WW^*\rightarrow 4j $& 5.34 & 2.85 & 152 & 9.03 & 3.77 & 3410 \\
        \hline
        $\mu^+\mu^-\rightarrow (\mu^+\mu^-,\nu_\mu\bar{\nu}_\mu)H; \ H\not\rightarrow WW^*$ & 422 & 0.48 & 2010 & 711 & 0.68 & 48500\\
        $\mu^+\mu^-\rightarrow (\mu^+\mu^-,\nu_\mu\bar{\nu}_\mu) jj$  & 1950 & 0.05 & 877 & 3020 & 0.09 & 28600\\
        $\mu^+\mu^-\rightarrow \nu_\mu\mu^\pm jj$  & 4340 & 0.02 & 948 & 7160 & 0.04 & 30200 \\ \hline
    \end{tabular}
    \caption{Signal and most important backgrounds for VBF $H\rightarrow WW^*\rightarrow 4j$ after applying flavor tagging and analysis cuts. $\mu^+\mu^-\rightarrow (\mu^+\mu^-,\nu_\mu\bar{\nu}_\mu)H; \ H\not\rightarrow WW^*$ includes all other hadronic modes and $\tau^+\tau^-$.}
    \label{tab:wwbackgrounds}
\end{table}

We study the $H\rightarrow WW^*$ decay mode in both the fully hadronic and semileptonic channels. The dominant backgrounds are different for each channel, with other hadronic $H$ decays contributing the most to the fully hadronic channel, while the requirement of an isolated lepton in the semileptonic channel suppresses other Higgs decays.

For the semileptonic channel, the signature is one isolated high energy lepton and two jets. Due to energy losses from the neutrino, the Higgs mass cannot be fully reconstructed, and likewise only one $W$ candidate can be reconstructed, which can be either on-shell or off-shell. For the fully hadronic channel, the signature is two pairs of light quark jets, one consistent with an on-shell $W$ decay, and a four-jet invariant mass consistent with a Higgs decay.

The semileptonic channel analysis begins by selecting events with two VLC jets with $R=0.5$ and one isolated lepton all with $p_T>20$ GeV. We select events with $5<m_{jj}<90$ GeV and $20<m_{jj\ell}<110$ GeV. We also impose cuts on the energy of the $W$ candidate and partially reconstructed Higgs depending on the collider energy. At 3 TeV, we impose $40<E_{jj}<700$ GeV and $85<E_H<800$ GeV. At 10 TeV, we impose $50<E_{jj}<1100$ GeV and $90<E_H<1600$ GeV. The invariant mass cuts remove most of the background events, especially those from on-shell diboson production. The energy cuts are necessary to further remove more of the dominant $\mu^+\mu^-\rightarrow (\nu_\ell\bar{\nu}_\ell,\ell^+\ell^-)\nu_\ell\ell^\pm jj$ background. The combination of these cuts yields a precision of 1.7\% at 3 TeV and 0.45\% at 10 TeV for the combined VBF mode. Incorporating forward muon tagging in addition to these cuts, we obtain a precision of 1.6(8.4)\% at 3 TeV and 0.42(2.0)\% at 10 TeV for $W^+W^-$($ZZ$) fusion.

In the fully hadronic channel, we first cluster events in exclusive mode into two jets, and remove events that have any $b$-tagged jets. We then cluster the event into four VLC jets with $R=0.5$ for analysis. We select only those events where all four jets have $p_T>40$ GeV. We then determine the combination of two jets that gives the closest value to the $W$ mass and assign it to the on-shell $W$-boson. The other two jets are assigned to the off-shell $W^*$. We then impose invariant mass cuts on both $W$ candidates and the Higgs candidate. We require $50<m_W<90$ for the on-shell $W$ candidate to remove on-shell $Z$ backgrounds, $15<m_{W^*}<50$ GeV for the off-shell $W^*$ and $100<m_H<135$ GeV on the Higgs candidate to remove on-shell diboson backgrounds. The most relevant remaining background events are summarised in Table \ref{tab:wwbackgrounds} and are dominated by $H\rightarrow (b\bar{b},c\bar{c},gg)$ being forced into four jets by exclusive clustering. The total VBF precision for this channel is found to be 5.7\% at 3 TeV and 1.3\% at 10 TeV. Using the same set of cuts while including forward muon tagging yields a precision of 5.4(17)\% at 3 TeV and 1.2(4.4)\% at 10 TeV for the $W^+W^-$($ZZ$) fusion mode.

\subsection{\texorpdfstring{$ZZ^*$}{ZZ*}}\label{wwf-zz}
\begin{table}[t]
    \centering
    \renewcommand{\arraystretch}{1.3}
    \setlength{\arrayrulewidth}{.3mm}
    \setlength{\tabcolsep}{0.5 em}
    \begin{tabular}{|c||c|c|c||c|c|c|}
    \multicolumn{1}{c}{} & \multicolumn{6}{c}{Number of Events} \\
    \hline
    \multirow{2}{*}{Process} & 
    \multicolumn{3}{c}{$3\,\textrm{TeV}$} & \multicolumn{3}{c|}{$10\,\textrm{TeV}$}\\ \cline{2-7}
    & $4j$ & $2j2\ell$ & $4\ell$ & $4j$ & $2j2\ell$ & $4\ell$ \\ \hline \hline    
        $\mu^+\mu^-\rightarrow \nu_\mu \bar{\nu}_\mu H; \ H\rightarrow ZZ^*\rightarrow X$ & 124 & 103 & 5 & 2910 & 1590 & 66 \\
        $\mu^+\mu^-\rightarrow \mu^+ \mu^- H; \ H\rightarrow ZZ^*\rightarrow X$  & 3 & 9 & 0 & 315 & 151 & 8 \\
        Backgrounds  & 6700 & 50 & 0 & 208000 & 1370 & 2 \\
        \hline
    \end{tabular}
    \caption{Total number of events for $WW/ZZ$-fusion $H\rightarrow ZZ^*\rightarrow X$ and combined backgrounds after all analysis cuts discussed in the text, where $X$ is one of $4j$, $2j2\ell$, or $4\ell$.}
    \label{tab:ZZ}
\end{table}

The $ZZ^*$ channel has three decay modes which we study- fully hadronic, semileptonic, and fully leptonic. In contrast with $WW^*$, the Higgs mass can be fully reconstructed for all three modes, at the cost of much lower signal statistics due to the lower $H\rightarrow ZZ^*$ branching fraction. 

In the fully hadronic decay mode, we impose very similar cuts to the fully hadronic $WW^*$ state. Events are clustered into four VLC jets with $R=0.5$, discarding events with isolated high-energy leptons. All four jets are required to have $p_{T,j}>40$ GeV. We then identify the combination of two jets yielding an invariant mass closest to $m_Z$, which are assigned to the on-shell $Z$. The other pair is assigned to the off-shell $Z^*$. We impose cuts on these invariant masses according to $15<m_{Z^*}<50$ GeV and $55<m_Z<95$ GeV. The reconstructed Higgs invariant mass is additionally required to lie in the range $100<m_H<135$ GeV. The combination of these mass cuts removes the majority of the contribution from single $W/Z$ decays. The remaining backgrounds then are dominated by two jet events that are clustered into four jets due to our exclusive clustering. In particular, the combined contribution of other single Higgs decays, especially $H\rightarrow b\bar{b}$ and $H\rightarrow WW^*\rightarrow 4j$, amounts to over 75\% of all the background events. While some of the $H\rightarrow b\bar{b}$ events can be removed with $b$-tagging information, this removes enough of the signal that it does not provide an increase in precision. The precision we find for the total VBF production mode in this channel is 65\% at 3 TeV and 14\% at 10 TeV. Considering forward tagging yields a precision of 65\% and 15\% for $W^+W^-$ fusion alone. Given the poor precision, we do not consider $ZZ$ fusion on its own.

The semileptonic decay mode is characterised by a pair of leptons and a pair of jets, with one pair's invariant mass consistent with an on-shell $Z$-boson. While this channel has substantially fewer signal events than the fully hadronic, the requirement of a lepton pair makes the channel relatively background-free. We require two leptons and two VLC jets with $R=0.5$, all with the looser cut of $p_T>20$ GeV. The pair of either leptons or jets with an invariant mass closer to $m_Z$ is assigned to the on-shell $Z$. We require the on-shell pair to have an invariant mass between $20<m_Z<100$ GeV and the off-shell pair to have an invariant mass between $5<m_{Z^*}<60$ GeV. We then apply different cuts on the total reconstructed Higgs invariant mass depending on which pair originates from the on-shell $Z$. If the lepton pair reconstructs the on-shell $Z$, we impose $100<m_H<130$ GeV. If the jet pair reconstructs the on-shell $Z$, we impose $80<m_H<135$ GeV, since the $m_H$ distribution is wider in this case. This combination of cuts removes the overwhelming majority of background events. The remaining events come primarily from $\mu^+\mu^-\rightarrow \nu_\ell\bar{\nu}_\ell\ell^+\ell^-jj$, with some smaller contributions from $\mu^+\mu^-\rightarrow \nu_\ell\ell^\pm\ell^+\ell^-jj$ and other Higgs decays. We find a precision of 11\% at 3 TeV and 3.2\% at 10 TeV for the total VBF production mode. With forward muon tagging information to distinguish the modes, we find a precision of 12(34)\% at 3 TeV and 3.4(11)\% at 10 TeV for $W^+W^-$($ZZ$) fusion.

The fully leptonic $ZZ^*$ decay mode, while the cleanest, suffers from low statistics. We require two pairs of leptons, all satisfying $p_T>20$, one pair from an on-shell $Z$, that combine to reconstruct the Higgs mass in the range $100<m_H<130$ GeV. The only non-negligible background contribution comes from $ZZ$ fusion $H\rightarrow ZZ^*$, contributing around 10\% as much as the $W^+W^-$ fusion signal. The resulting precision in this channel from the total VBF production mode is found to be 45\% at 3 TeV and 12\% at 10 TeV. Removing $ZZ$ fusion contributions with forward tagging yields 48\% at 3 TeV and 13\% at 10 TeV for $W^+W^-$ fusion. As with $ZZ(4j)$, we do not consider $ZZ$ fusion on its own given the poor statistics of the channel. A summary of the number of events for all three channels is shown in Table \ref{tab:ZZ}.

\subsection{\texorpdfstring{$\gamma\gamma$}{γγ}}\label{wwf-gammagamma}
\begin{table}[t]
    \centering
    \renewcommand{\arraystretch}{1.3}
    \setlength{\arrayrulewidth}{.3mm}
    \setlength{\tabcolsep}{0.8 em}
    \begin{tabular}{|c||c|c|c||c|c|c|}
    \hline
    \multirow{2}{*}{Process} & 
    \multicolumn{3}{c}{$3\,\textrm{TeV}$} & \multicolumn{3}{c|}{$10\,\textrm{TeV}$}\\ \cline{2-7}
    & $\sigma \ (\textrm{fb})$ & $\epsilon \ (\%)$ & N & $\sigma \ (\textrm{fb})$ & $\epsilon\  (\%)$ & N \\ \hline \hline    
        $\mu^+\mu^-\rightarrow \nu_\mu \bar{\nu}_\mu H; \ H\rightarrow \gamma\gamma$ & 1.14 & 37 & 419 & 1.92 & 29 & 5550 \\
        $\mu^+\mu^-\rightarrow\mu^+\mu^- H; \ H\rightarrow \gamma\gamma$ & 0.12 & 29 & 34 & 0.20 & 29 & 576 \\
        \hline
        $\mu^+\mu^-\rightarrow \nu_\mu\bar{\nu}_\mu\gamma\gamma$ & 198 & 0.16 & 311 & 378 & 0.10 & 3750  \\
        $\mu^+\mu^-\rightarrow \mu^+\mu^-\gamma\gamma$ & 297 & 0.001 & 3 & 307 & 0.0002 & 7  \\\hline
    \end{tabular}
    \caption{Signal and considered backgrounds for the $H\rightarrow \gamma\gamma$ channel, after applying the cuts listed in the text.}
    \label{tab:aabackgrounds}
\end{table}

\begin{figure}[t]
    \centering
    \includegraphics[width=.49\textwidth]{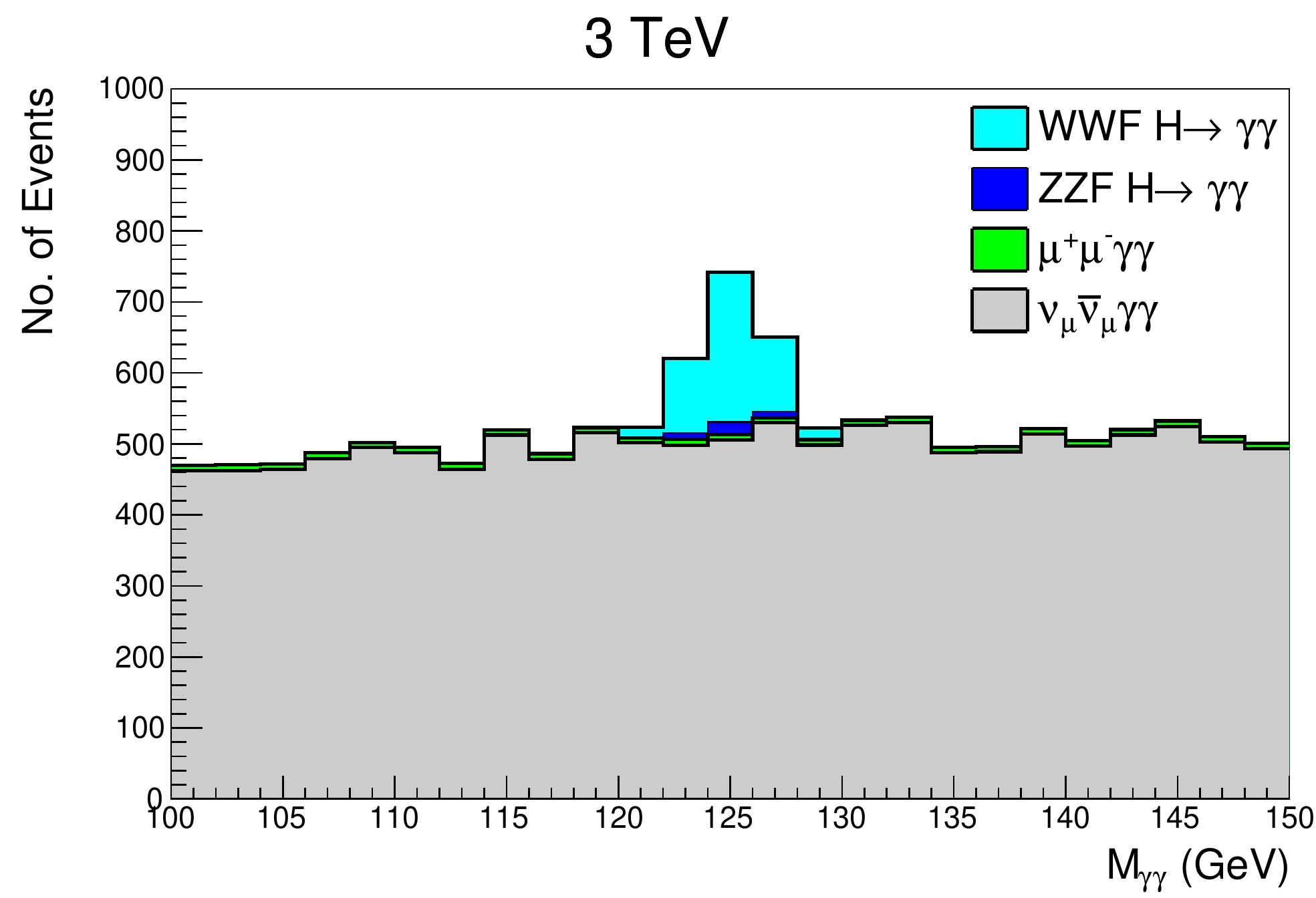}
    \includegraphics[width=.49\textwidth]{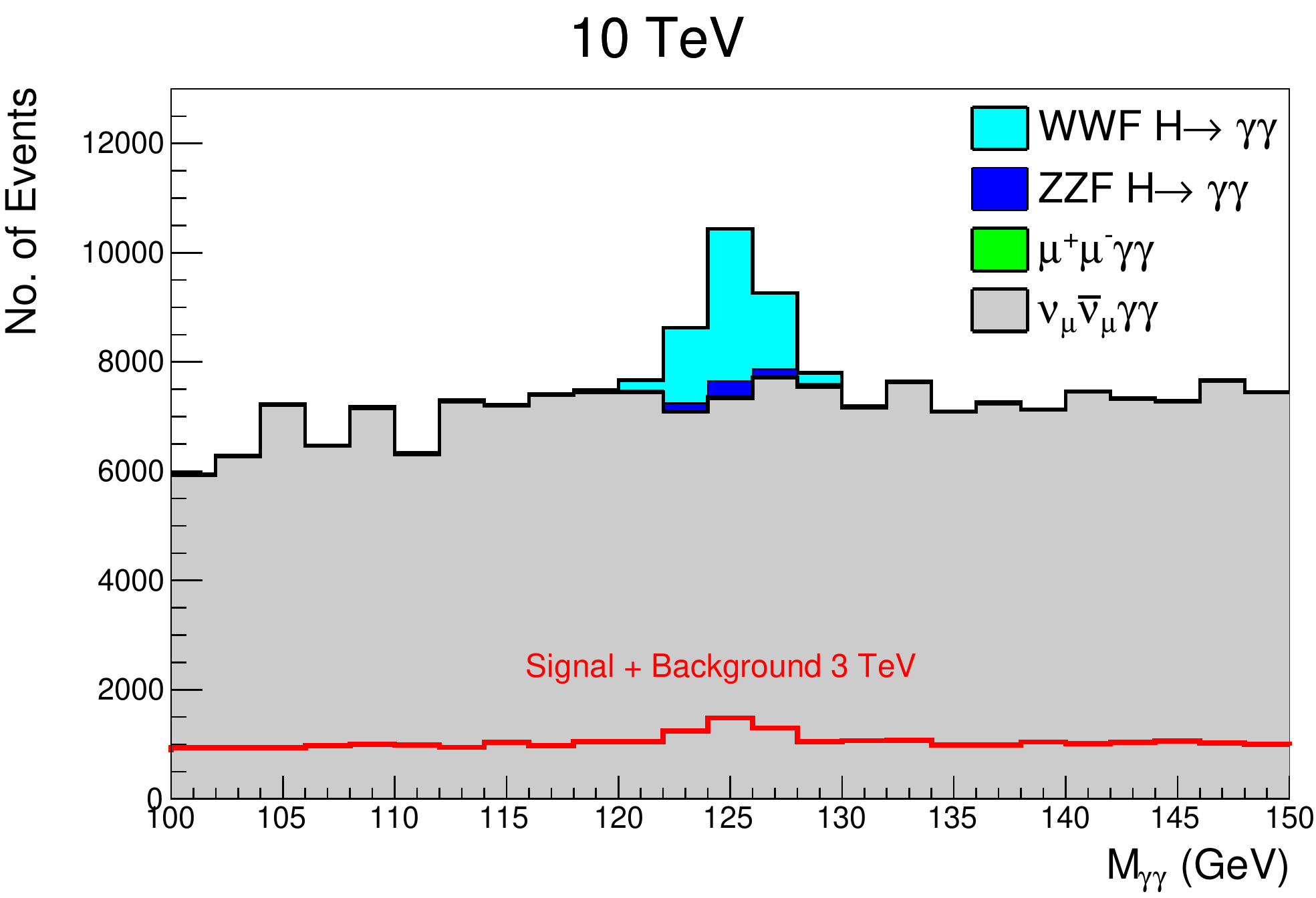}
    \caption{Stacked $\gamma\gamma$ pair invariant mass histograms for the $H\rightarrow \gamma\gamma$ analysis after cuts at (left) 3 TeV and (right) 10 TeV. The sum of signal and background at 3 TeV is overlayed on the 10 TeV for ease of comparison.}
    \label{fig:aaHists}
\end{figure}

For the $H\rightarrow \gamma\gamma$ channel, we select events with at least two isolated photons and no jets or leptons. The highest two $p_T$ photons are required to have $p_T>40$ GeV and are identified with the Higgs decay products. The resulting invariant mass distributions at 3 and 10 TeV are shown in Figure \ref{fig:aaHists}. We then impose a simple invariant mass cut on the Higgs candidate of $122<m_H<128$ GeV which removes the vast majority of the continuum background. The remaining background after these cuts are summarised in Table \ref{tab:aabackgrounds}, where we can see that $\nu_\mu\bar{\nu}_\mu\gamma\gamma$ contributes the vast majority of background events. We find a precision of 6.1\% at 3 TeV and 1.6\% at 10 TeV for the combined VBF production mode. If we include forward muon tagging information, we obtain results of 6.4(23)\% at 3 TeV and 1.7(4.8)\% at 10 TeV for $W^+W^-$($ZZ$) fusion.

\subsection{\texorpdfstring{$Z\gamma$}{Zγ}}\label{wwf-zgamma}
\begin{table}[t]
    \centering
    \renewcommand{\arraystretch}{1.3}
    \setlength{\arrayrulewidth}{.3mm}
    \setlength{\tabcolsep}{0.68 em}
    \begin{tabular}{|c||c|c|c||c|c|c|}
    \hline
    \multirow{2}{*}{Process} & 
    \multicolumn{3}{c}{$3\,\textrm{TeV}$} & \multicolumn{3}{c|}{$10\,\textrm{TeV}$}\\ \cline{2-7}
    & $\sigma \ (\textrm{fb})$ & $\epsilon \ (\%)$ & N & $\sigma \ (\textrm{fb})$ & $\epsilon\  (\%)$ & N \\ \hline \hline    
        $\mu^+\mu^-\rightarrow \nu_\mu \bar{\nu}_\mu H; \ H\rightarrow Z\gamma$ & 0.77 & 7.4 & 57 & 1.29 & 6.2 & 805 \\
        $\mu^+\mu^-\rightarrow \mu^+ \mu^- H; \ H\rightarrow Z\gamma$ &  0.08 &  5.6 & 4 & 0.13 & 6.6 & 88 \\
        \hline
        $\mu^+\mu^-\rightarrow (\nu_\mu \bar{\nu}_\mu,\mu^+\mu^-) H; \  H \not\rightarrow Z\gamma $ & 548 & 0.040 & 222 & 931 & 0.036 & 3320 \\
        $\mu^+\mu^-\rightarrow (\nu_\mu \bar{\nu}_\mu,\mu^+\mu^-) jj$ & 1950 & 0.002 & 40 & 3020 & 0.002 & 503 \\
        $\mu^+\mu^-\rightarrow \nu_\mu\mu^\pm jj$ & 4340 & 0.001 & 56 & 7160 & 0.002 & 1220\\
        $\mu^+\mu^-\rightarrow \nu_\mu \bar{\nu}_\mu jj\gamma$ & 75.6 & 0.45 & 342 & 161 & 0.25 & 4020\\ 
        $\mu^+\mu^-\rightarrow \nu_\mu\mu^\pm jj\gamma$ & 123 & 0.092 & 114 & 209 & 0.16 & 3260 \\
        \hline
    \end{tabular}
    \caption{Signal and most important backgrounds for the hadronic $H\rightarrow Z\gamma$ decay mode, after applying the cuts listed in the text. Processes without an explicit $\gamma$ have had the photon generated from showering.}
    \label{tab:zgambackgrounds}
\end{table}

For $H\rightarrow Z\gamma$, we only study the dominant hadronic decay mode of the $Z$. After event generation, we use a modified version of the {\sc Delphes} card to separate out the highest $p_T$ photon in the event and cluster the remaining particles into two $R=1.2$ Valencia jets. Events with isolated muons are removed to suppress $ZZ/WZ$ fusion backgrounds. We then apply cuts on the transverse momenta of the jets and the hard photon of $p_{T,j}>20$ GeV and $p_{T,\gamma}>40$ GeV respectively, harder on the photon to remove soft FSR backgrounds from showering. An additional cut on the distance between the hard photon and both jets of $\Delta R_{\gamma,j}>0.15$ is applied to remove backgrounds with collinear photons from showering. The invariant mass of the two jets is required to be in the range $85<m_{jj}<100$ GeV to be consistent with a $Z$ decay, with a tighter lower bound to suppress backgrounds from $W$ decays. The total invariant mass of the Higgs candidate is additionally required to lie in the range $105<m_H<130$ GeV. The signal and most important backgrounds are summarised in Table \ref{tab:zgambackgrounds}. Processes without an explicit $\gamma$ have had the photon generated from showering. We find a precision of 47\% at 3 TeV and 13\% at 10 TeV for the combined VBF signal. If we additionally allow for forward muon tagging, we can improve the sensitivity of $W^+W^-$ fusion to 45\% at 3 TeV and 12\% at 10 TeV. The $ZZ$ fusion mode's contribution is so small that we do not consider it seperately.

\subsection{\texorpdfstring{$\mu^+\mu^-$}{μμ}}\label{wwf-mumu}

The $H\rightarrow \mu^+\mu^-$ decay channel is challenging due to the small branching ratio of 0.22\%. The backgrounds are dominated by the continuum of $\nu_\mu\bar{\nu}_\mu\mu^+\mu^-$ and $4\mu$, both of which arise primarily from VBF subprocesses at high energies. We analyse events with two oppositely charged muons with  $p_T>20$ GeV and $|\eta|<2.5$, where we have loosened the $p_T$ cut compared to other $2\rightarrow2$ processes. We note that we get results that are only slightly worse if the $p_T$ cut is increased to 30 GeV. We impose an invariant mass cut on the Higgs candidate of $124<m_H<126$ GeV to remove the majority of the continuum background. 

A summary of the signal and relevant backgrounds is included in Table \ref{tab:mumubackgrounds}. We can see that the $4\mu$ contribution is significantly larger at 10 TeV than other backgrounds. We find a precision of 40\% at 3 TeV and 9.8\% at 10 TeV for the combined VBF signal. Including forward muon tagging, we can reduce the dominant $4\mu$ background significantly for the $W^+W^-$ fusion signal, yielding a precision of 28\% at 3 TeV and 5.8\% at 10 TeV.

\begin{table}[t]
    \centering
    \renewcommand{\arraystretch}{1.3}
    \setlength{\arrayrulewidth}{.3mm}
    \setlength{\tabcolsep}{0.8 em}
    \begin{tabular}{|c||c|c|c||c|c|c|}
    \hline
    \multirow{2}{*}{Process} & 
    \multicolumn{3}{c}{$3\,\textrm{TeV}$} & \multicolumn{3}{c|}{$10\,\textrm{TeV}$}\\ \cline{2-7}
    & $\sigma \ (\textrm{fb})$ & $\epsilon \ (\%)$ & N & $\sigma \ (\textrm{fb})$ & $\epsilon\  (\%)$ & N \\ \hline \hline    
        $\mu^+\mu^-\rightarrow \nu_\mu \bar{\nu}_\mu H; \ H\rightarrow \mu^+\mu^-$ & 0.11 & 52 & 57 & 0.18 & 39 & 715  \\
        $\mu^+\mu^-\rightarrow \mu^+ \mu^- H; \ H\rightarrow \mu^+\mu^-$ & 0.011 & 43 & 5 & 0.019 & 39 & 73  \\
        \hline
        $\mu^+\mu^-\rightarrow \nu_\mu \bar{\nu}_\mu \mu^+\mu^-$ & 67.2 & 0.30 & 198 & 71.5 & 0.13 & 960 \\
        $\mu^+\mu^-\rightarrow \mu^+\mu^- \mu^+\mu^-$ & 211 & 0.16 & 345 & 307 & 0.14 & 4191 \\ \hline
    \end{tabular}
    \caption{Signal and most important backgrounds for the $H\rightarrow \mu^+\mu^-$ channel, after applying the cuts listed in the text.}
    \label{tab:mumubackgrounds}
\end{table}

\subsection{\texorpdfstring{$t\bar{t}H$}{ttH}}\label{ttH}
In principle, at a high energy muon collider, the top quark Yukawa coupling can be accessed in a variety of indirect ways.  However, direct $t\bar{t}H$ production allows for, in some sense, the most direct testing of it. The cross section for $s$-channel $t\bar{t}H$ production falls off at the high energies we are interested in, and in particular gets surpassed by VBF $t\bar{t}H$ production at $\sim7$ TeV, as discussed in Section \ref{sec:methodology}. The analysis for 3 TeV $t\bar{t}H$ production is therefore dominated by $s$-channel production, while at 10 TeV it is dominated by VBF production, which have very different kinematic features and so must be considered independently. Only the $H\rightarrow b\bar{b}$ decay mode is considered, since other channels are unlikely to have large enough statistics to be viable.

We start with 3 TeV. The signature for $t\bar{t}H$ production depends on the decay modes of the top quarks. We therefore consider two cases. First, the fully hadronic case, where both top quarks decay as $t\rightarrow b W\rightarrow bjj$, resulting in a total of 8 final state jets, four of which are $b$-jets. The second case is where one $W$-boson decays semileptonically, in which case the final state is one lepton and 6 jets- four $b$ and two light. We handle these two decay modes independently.

We first look at each event for any isolated leptons. If we see a single isolated lepton, we classify the event is semileptonic. If we see none, we classify it as fully hadronic. Events with more than one isolated lepton are discarded. Events are then clustered inclusively into VLC jets with $R=0.5$. Any event with less than four jets from this clustering is thrown away. We then recluster the event using the exclusive VLC algorithm. If the event is semileptonic, we cluster the event into 6 jets with $R=1.5$. If the event is fully hadronic, we cluster it into 8 jets with $R=0.5$. All of these jets are required to have $p_T>20$ GeV or the event is discarded.

After clustering, we search for $b$-tagged jets. We require at least two tight $b$-tags (50\% working point). If we find fewer than 4 tight $b$-tags, we search for the highest $p_T$ loose $b$-tags (90\% working point) until we have four $b$-tags in total; if we do not find four, the event is discarded. Kept events have their remaining 2 (semileptonic channel) or 4 (hadronic channel) jets classified as light jets originating from $W$-decay.

We then assign jets to parent particles in the same way as done by CLIC \cite{Abramowicz:2016zbo}. For the semileptonic and hadronic channels we design separate likelihood functions that are minimized by the correct pairing of jets.  After assigning jets to parents, in the semileptonic channel, we impose the series of cuts listed below:
\begin{multicols}{2}
\begin{itemize}
    \item $E_H>50$ GeV
    \item $E_{W_{1,2}}>550$ GeV
    \item $m_H>50$ GeV
    \item $m_{t_{1,2}}>90$ GeV
    \item $\theta_{t_{1,2},H}>2^\circ$
    \item[\vspace{\fill}]
\end{itemize}
\end{multicols}
The combination of these cuts is meant primarily to minimize the impact of the dominant $t\bar{t}$ background, where some of the jets are split into multiple by our exclusive jet clustering. In the hadronic channel, we find that there are few cuts that are helpful in distinguishing the signal and this background, and only impose a cut $E_{W_{1,2}}>400$ GeV.

We find a precision of 81\% in the fully hadronic case and 88\% in the semileptonic case. Combined, we find a precision of $61\%$ for $s$-channel $t\bar{t}H$ production at 3 TeV. The backgrounds are heavily dominated by $t\bar{t}$, where $b$-jets are split into multiple jets by our exclusive clustering, and our flat $b$-tagging identified them as separate $b$-jets.

At 10 TeV, we follow a similar set of steps as in 3 TeV, although our cuts are quite different owing to the different kinematics of vector boson fusion and $s$-channel production. After identification of semileptonic or fully hadronic, we inclusively cluster with $R=0.5$ and set cuts on the number of jets from this clustering of $N_j>4(5)$ for the semileptonic(hadronic) channel. We then proceed to force the event into 6(8) jets with $R=1.5(0.5)$. Following the same procedure as in the 3 TeV case for $b$-tagging, we select events with four $b$-jets and two(four) light jets passing $p_T>20$ GeV cut. 

Particles are again identified with parent particles using likelihood functions. For the semileptonic channel, we impose cuts of $E_H>120$ GeV and $E_{W_{1,2}}>180$ GeV. Additionally, we impose a cut on the total detected invariant mass of the six jets of $450<m_{Total}<4000$ GeV, which eliminates $s$-channel backgrounds and some of the VBF $t\bar{t}$ background. In the hadronic channel, we only apply a cut on the total mass of the eight jet system of $600<m_{Total}<5000$ GeV.

\begin{table}[t]
    \centering
    \renewcommand{\arraystretch}{1.3}
    \setlength{\arrayrulewidth}{.3mm}
    \setlength{\tabcolsep}{0.6 em}
    \begin{tabular}{|c||c|c|c|c|}
    \multicolumn{1}{c}{} & \multicolumn{4}{c}{Number of Events} \\
    \hline
    \multirow{2}{*}{Process} & \multicolumn{2}{c}{$3\,\textrm{TeV}$} & \multicolumn{2}{c|}{$10\,\textrm{TeV}$}\\\cline{2-5}
    & SL & Had & SL & Had \\\cline{2-5} \hline \hline    
        $t\bar{t}H; \ H\rightarrow b\bar{b}$ & 34 & 63 & 49 & 59 \\
        \hline
        $t\bar{t}H; \ H \not\rightarrow b\bar{b}$ & 9 & 21 & 6 & 11 \\
        $t\bar{t}$ & 609 & 2070 & 502 & 1440 \\
        $t\bar{t}Z $ & 207 & 362 & 530 & 663 \\
        $t\bar{t} b\bar{b}$ & 9 & 21 & 15 & 18 \\
        \hline
    \end{tabular}
    \caption{A summary of the signal and background events for $t\bar{t}H$ production in the $H\rightarrow b\bar{b}$ decay channel after all analysis cuts. All of these processes include both $s$-channel production and VBF production modes.}
    \label{tab:tth}
\end{table}

As in the 3 TeV case, $t\bar{t}$ contributes the majority of the background in the fully hadronic case. $t\bar{t}Z$ also contributes a sizable amount of background, even more than $t\bar{t}$ in the SL case. We find a precision of 79\% in the fully hadronic case and 68\% in the semileptonic case, for a combined precision of 53\% at 10 TeV. A summary of the signal and backgrounds considered for $t\bar{t}H$ at both 3 and 10 TeV can be seen in \ref{tab:tth}. We emphasize that while this channel in particular can be improved with more complex flavor tagging techniques to distinguish between the four $b$-jet $t\bar{t}H$ and the two $b$-jet $t\bar{t}$, in general the cross section for $t\bar{t}H$ is so small at these energies that it is unlikely to give a measurement competitive with $pp$ or $e^+e^-$ colliders beyond the $t\bar{t}H$ threshold even after improvement.

\begin{table}[t]
\renewcommand{\arraystretch}{1.3}
\setlength{\arrayrulewidth}{.3mm}
\setlength{\tabcolsep}{0.8 em}
\begin{center}
\begin{tabular}{|c|c|c|} %\hline
    \hline
    \multirow{2}{*}{Channel} & 
    \multicolumn{2}{c|}{$\Delta\sigma/\sigma$ (\%)}\\ \cline{2-3}
    & $3\,\textrm{TeV}$ & $10\,\textrm{TeV}$ \\ \hline \hline
$bb$ & $0.76$ & $0.21$ \\
$cc$ & 13 & 4.0 \\
$gg$ & 3.3 & 0.89 \\
$\tau^+\tau^-$ & 4.0 & 1.1 \\
$WW^*(jj\ell \nu)$ & 1.7 & 0.45 \\
$WW^*(4j)$ & 5.7 & 1.3 \\
$ZZ^*(4\ell)$ & 45 & 12 \\
$ZZ^*(jj\ell\ell)$ & 11 & 3.2 \\
$ZZ^*(4j)$ & 65 & 14 \\
$\gamma\gamma$ & $6.1$ & $1.6$ \\
$Z(jj)\gamma$ & 47 & 13 \\
$\mu^+\mu^-$ & $40$ & $9.8$ \\
\hline
$ttH(bb)$ & 61 & 53 \\ \hline
\end{tabular}
\end{center}
\caption{Signal precision for selected Higgs production channels at $3\,\textrm{TeV}$ and $10\,\textrm{TeV}$ muon colliders using {\sc Delphes} fast simulation including physics backgrounds. All channels are for the combined $W^+W^-$ fusion and $ZZ$ fusion processes. $t\bar{t}H$ includes all of $s$-channel, $ZZ$ fusion, and $W^+W^-$ fusion production modes.
}\label{tab:higgsrateprecision}
\end{table}

\begin{table}[t]
\renewcommand{\arraystretch}{1.3}
\setlength{\arrayrulewidth}{.3mm}
\setlength{\tabcolsep}{0.8 em}
\begin{center}
\begin{tabular}{|c|c|c|c||c|} %\hline
    \hline
    \multirow{2}{*}{Production} & 
    \multirow{2}{*}{Decay} & 
    \multicolumn{2}{c||}{$\Delta\sigma/\sigma$ (\%)} & Signal Only \\ \cline{3-5}
    && $3\,\textrm{TeV}$ & $10\,\textrm{TeV}$ & $10\,\textrm{TeV}$ \\ \hline \hline
\multirow{12}{*}{$W^+W^-$ fusion} & $bb$ & 0.80 & 0.22 & 0.17 \\ \cline{2-5}
& $cc$ & 12 & 3.6 & 1.7\\ \cline{2-5}
& $gg$ & 2.8 & 0.79 & 0.19\\ \cline{2-5}
& $\tau^+\tau^-$ & 3.8 & 1.1 & 0.54\\ \cline{2-5}
& $WW^*(jj\ell \nu)$ & 1.6 & 0.42 & 0.30\\ \cline{2-5}
& $WW^*(4j)$ & 5.4 & 1.2 & 0.49\\ \cline{2-5}
& $ZZ^*(4\ell)$ & 48 & 13 & 12\\ \cline{2-5}
& $ZZ^*(jj\ell\ell)$ & 12 & 3.4 & 2.3\\ \cline{2-5}
& $ZZ^*(4j)$ & 65 & 15 & 1.4\\ \cline{2-5}
& $\gamma\gamma$ & 6.4 & 1.7 & 1.3\\ \cline{2-5}
& $Z(jj)\gamma$ & 45 & 12 & 2.0\\ \cline{2-5}
& $\mu^+\mu^-$ & 28 & 5.7 & 3.9\\ \cline{2-5}
\hline
\multirow{8}{*}{$ZZ$ fusion} & $bb$ & 2.6 & 0.77 & 0.49\\ \cline{2-5}
& $cc$ & 72 & 17 & -\\ \cline{2-5}
& $gg$ & 14 & 3.3 & -\\ \cline{2-5}
& $\tau^+\tau^-$ & 21 & 4.8 & -\\ \cline{2-5}
& $WW^*(jj\ell \nu)$ & 8.4 & 2.0 & -\\ \cline{2-5}
& $WW^*(4j)$ & 17 & 4.4 & 1.3\\ \cline{2-5}
& $ZZ^*(jj\ell\ell)$ & 34 & 11 & -\\ \cline{2-5}
& $\gamma\gamma$ & 23 & 4.8 & -\\ \cline{2-5}
\hline
$ttH$ & $bb$ & 61 & 53 & 12\\ \hline
\end{tabular}
\end{center}
\caption{Signal precision for selected Higgs production channels at $3\,\textrm{TeV}$ and $10\,\textrm{TeV}$ muon colliders using {\sc Delphes} fast simulation including physics backgrounds with a hypothetical forward muon detector up to $|\eta|<6$. The signal-only results from~\cite{SmashersGuide} are shown for comparison, with a typo in the $\mu^+\mu^-$ channel corrected. $t\bar{t}H$ includes all of $s$-channel, $ZZ$ fusion, and $W^+W^-$ fusion production modes.
}\label{tab:higgsrateprecisionFwdTagging}
\end{table}

\section{Higgs coupling precision}\label{sec:couplings}

Before moving on to Higgs coupling precisions, we summarize our overall results from the previous section for cross section sensitivities for each channel in Table \ref{tab:higgsrateprecision} without forward muon tagging and Table \ref{tab:higgsrateprecisionFwdTagging} including it. Table \ref{tab:higgsrateprecisionFwdTagging} additionally shows the 10 TeV signal-only results from~\cite{SmashersGuide} for comparison. All channels for the former combine $W^+W^-$ fusion and $ZZ$ fusion as the signal. In the latter, we have used the forward muon tagging up to $|\eta|<6$ to count the number of extra muons in the event and required either $N_\mu=0$ or $N_\mu=2$, without any further cuts or optimisation.

It is useful to briefly comment on some of the differences in Tables~\ref{tab:higgsrateprecision} and~\ref{tab:higgsrateprecisionFwdTagging}. Without forward tagging, it is impossible to distinguish between $W^+W^-$ and $ZZ$ fusion, making any statements about one or the other alone impossible. This directly translates into worse precision on both of the Higgs couplings to the $W^\pm$ and the $Z$, as we will see. Additionally, backgrounds from all $\mu^+\mu^-\rightarrow (\nu_\mu\bar{\nu}_\mu,\mu_\nu\mu^\pm,\mu^+\mu^-)X$ backgrounds contribute, while forward tagging makes the backgrounds for $W^+W^-$ and $ZZ$ fusion different, heavily suppressing $(\nu_\mu\mu^\pm)X$ processes for both. This is particularly impactful for $H\rightarrow gg$, where tagging information cannot be used to remove backgrounds from $W$ decays. Generally speaking, the precision without forward tagging is similar to the precision in only $W^+W^-$ fusion with it.

While the results shown in Tables~\ref{tab:higgsrateprecision} and~\ref{tab:higgsrateprecisionFwdTagging} contain the important information about the sensitivities in the various single Higgs channels, it's a useful exercise to understand what this implies for measuring the Higgs couplings themselves.  The translation from ``experimental'' measurements to interpretations is, of course, fraught with difficulties, as there are always caveats for when a given framework breaks down in an attempt to do things in a semi-model independent way.  The standard approaches to Higgs coupling precision fits are the ``kappa'' framework and the EFT approach~\cite{deBlas:2019rxi}.  The EFT approach is particularly useful in terms of the symmetries it preserves, but it often introduces degeneracies that require additional non-Higgs observables so that the fits look sensible, even if it is not necessary in broad classes of models.  To demonstrate solely the effects of Higgs precision, we use as a first estimate the ``kappa'' framework as discussed in detail in~\cite{LHC-HXSWG:2012,LHC-HXSWG:2013}.  Furthermore, since there is not a study which measures the Higgs width solely with a high energy muon collider yet, we restrict ourselves to the kappa-0 framework, where no new BSM decay modes are included in the total Higgs width~\cite{deBlas:2019rxi}.  For the interested reader, our results for cross section sensitivities have been included in an EFT fit shown in the recent IMCC white paper~\cite{DeBlas:2022wxr}.

\begin{table}[t]
\renewcommand{\arraystretch}{1.5}
\setlength{\arrayrulewidth}{.3mm}
\setlength{\tabcolsep}{1.1 em}
\begin{center}
\begin{tabular}{|c|cc|cc|cc|}
\multicolumn{7}{c}{Fit Result [\%]} \\ \hline
 & \multicolumn{2}{c|}{$\mu^+\mu^-$} & \multicolumn{2}{c|}{+ HL-LHC} & \multicolumn{2}{c|}{+ HL-LHC + $250\,\textrm{GeV}$ $e^+e^-$ } \\ 
 & 3 TeV & 10 TeV & 3 TeV & 10 TeV & 3 TeV & 10 TeV\\ \hline
$\kappa_W$ & 0.55 & 0.16 & 0.39 & 0.14 & 0.33 & 0.11\\ 
$\kappa_Z$ & 5.1 & 1.4 & 1.3 & 0.94 & 0.12 & 0.11\\ 
$\kappa_g$ & 2.0 & 0.52 & 1.4 & 0.50 & 0.75 & 0.43\\ 
$\kappa_{\gamma}$ & 3.2 & 0.84 & 1.3 & 0.71 & 1.2 & 0.69\\ 
$\kappa_{Z\gamma}$ & 24 & 6.5 & 24 & 6.5 & 4.1 & 3.5\\ 
$\kappa_c$ & 6.8 & 2.0 & 6.7 & 2.0 & 1.8 & 1.3\\ 
$\kappa_t$ & 35 & 55 & 3.2 & 3.2 & 3.2 & 3.2\\ 
$\kappa_b$ & 0.97 & 0.26 & 0.82 & 0.25 & 0.45 & 0.22\\ 
$\kappa_{\mu}$ & 20 & 4.9 & 4.6 & 3.4 & 4.1 & 3.2\\ 
$\kappa_{\tau}$ & 2.3 & 0.63 & 1.2 & 0.57 & 0.62 & 0.41 \\\hline
\end{tabular}
\end{center}
\caption{
Results of a 10-parameter fit to the Higgs couplings in the $\kappa$-framework, based on the attainable precision in each production and decay channel without forward muon tagging listed in Table~\ref{tab:higgsrateprecision}.}\label{tab:higgscouplingfitNoFwd}
\end{table}

For completeness, we define coupling modifiers $\kappa_i$ as the ratio between the measured and the standard model value, such that in the narrow-width approximation, for the production mode $i$ and Higgs decay channel $f$, we have
\begin{equation}
    \mu_i = \frac{\sigma_i\cdot\text{BR}_f}{\sigma_{i}^{\text{SM}}\cdot\text{BR}_{f}^{\text{SM}}} = \frac{\kappa_i^2\kappa_f^2}{\kappa_H^2} , \qquad \qquad \kappa_H = \sum_f \frac{\kappa_f^2\Gamma_f^{SM}}{\Gamma_H^{SM}}
\end{equation}
where $\Gamma_f$ is the partial width for the channel $f$. In this way, the Higgs couplings are set back to their SM values in the limit where all $\kappa_i \rightarrow 1$. For the loop induced processes $H\rightarrow (\gamma\gamma,gg,Z\gamma)$, we introduce effective couplings $\kappa_\gamma$, $\kappa_g$, and $\kappa_{Z\gamma}$ for a total of 10 fitting parameters. For the case without forward tagging, since we cannot seperately identify $W^+W^-$ and $ZZ$ fusion, we instead use signal strengths of $\mu_i = (\frac{N^{WWF}}{N^{Total}})\mu^{WWF}_{i}+(\frac{N^{ZZF}}{N^{Total}})\mu^{ZZF}_{i}$ in the fit, where $N^{WWF}$($N^{ZZF}$) is the number of events contributed by the $W^+W^-$($ZZ$) fusion production mode and $N^{Total} = N^{WWF}+N^{ZZF}$.

For $t\bar{t}H$, there are multiple production modes that enter into the cross section, so it must be handled with care. In particular, in addition to the contribution from the top Yukawa coupling, there are contributions from the Higgs radiating off of a $s$-channel $Z$ boson or $t$-channel $Z/W$ boson. We therefore treat this channel somewhat differently from the others. Since the precision on $\sigma_{t\bar{t}H}$ attained in Section \ref{ttH} is so much worse than those for $W^+W^-$ fusion $H\rightarrow WW^*/ZZ^*$, the precision on $\kappa_{Z,W}$ should not be affected much by this channel, and likewise the precision in $\kappa_t$ should not be significantly impacted by their precision. For simplicity, we therefore fix them to the SM. We then compute the total VBF+$s$-channel cross section as a function of $y_t$ from $0.25\leq y_t/y_t^{SM} \leq 1.75$ at each energy and express the signal strength of the channel as $\mu_{t\bar{t}H}=(a \kappa_t^2 + b \kappa_t + c)\kappa_b^2/(\kappa_H^2\sigma^{SM})$. For $s$-channel, this reproduces a very similar rescaling to that of CLIC \cite{Abramowicz:2016zbo}, while for VBF we keep the contributions from interference terms that are crucial in maintaining perturbative unitarity.

\begin{table}[t]
\renewcommand{\arraystretch}{1.5}
\setlength{\arrayrulewidth}{.3mm}
\setlength{\tabcolsep}{1.1 em}
\begin{center}
\begin{tabular}{|c|cc|cc|cc|}
\multicolumn{7}{c}{Fit Result with Forward Muon Tagging [\%]} \\ \hline
 & \multicolumn{2}{c|}{$\mu^+\mu^-$} & \multicolumn{2}{c|}{+ HL-LHC} & \multicolumn{2}{c|}{+ HL-LHC + $250\,\textrm{GeV}$ $e^+e^-$ } \\ 
 & 3 TeV & 10 TeV & 3 TeV & 10 TeV & 3 TeV & 10 TeV \\ \hline
$\kappa_W$ & 0.37 & 0.10 & 0.35 & 0.10 & 0.31 & 0.10 \\ 
$\kappa_Z$ & 1.2 & 0.34 & 0.89 & 0.33 & 0.12 & 0.11 \\
$\kappa_g$ & 1.6 & 0.45 & 1.3 & 0.44 & 0.72 & 0.39 \\
$\kappa_{\gamma}$ & 3.2 & 0.84 & 1.3 & 0.71 & 1.2 & 0.69 \\
$\kappa_{Z\gamma}$ & 21 & 5.5 & 22 & 5.5 & 4.0 & 3.3 \\
$\kappa_c$ & 5.8 & 1.8 & 5.8 & 1.8 & 1.7 & 1.3 \\
$\kappa_t$ & 34 & 53 & 3.2 & 3.2 & 3.2 & 3.2 \\
$\kappa_b$ & 0.84 & 0.23 & 0.80 & 0.23 & 0.44 & 0.21 \\
$\kappa_{\mu}$ & 14 & 2.9 & 4.7 & 2.5 & 4.0 & 2.4 \\
$\kappa_{\tau}$ & 2.1 & 0.59 & 1.2 & 0.55 & 0.61 & 0.40 \\ \hline
\end{tabular}
\end{center}
\caption{
Results of a 10-parameter fit to the Higgs couplings in the $\kappa$-framework, based on the attainable precision in each production and decay channel with forward muon tagging up to $|\eta|<6$ listed in Table~\ref{tab:higgsrateprecisionFwdTagging}.}\label{tab:higgscouplingfitWithFwd}
\end{table}

The results of this fit using the data without a forward muon detector in Table \ref{tab:higgsrateprecision} are shown in Table \ref{tab:higgscouplingfitNoFwd}.  In addition, we show results where we perform this fit in combination with the HL-LHC and with a 250 GeV $e^+e^-$ collider in columns 2 and 3 of these tables. For the HL-LHC input, we use the CMS results and correlation matrix with S2 systematics\footnote{The HL-LHC inputs are taken from the publicly available results at \url{https://twiki.cern.ch/twiki/bin/view/LHCPhysics/GuidelinesCouplingProjections2018}.} \cite{Cepeda:2019klc}. For the 250 GeV $e^+e^-$ collider input, we use the results from CEPC with full correlation matrix \cite{CEPCStudyGroup:2018ghi,An:2018dwb}. Results in combination with other $e^+e^-$ colliders such as the ILC  or FCC-ee would be similar. We also perform this fit in the scenario where we have muon tagging capabilities up to $|\eta|<6$ in Table \ref{tab:higgscouplingfitWithFwd}. Most of the channels improve, with $\kappa_Z$ improving the most. This makes sense, given how challenging the $H\rightarrow ZZ^*$ decay channels are to measure, either due to low statistics in the leptonic channels or high backgrounds in the hadronic channels.

\section{Conclusion and Future Work}\label{sec:conclusions}

We have presented the first analysis of the potential sensitivity of future muon colliders to most relevant production and decay modes using {\sc Delphes} fast simulation at 3 TeV at 1/ab and 10 TeV at 10/ab while including physics backgrounds. Using a simple cut-and-count analysis strategy, we have demonstrated that a 10 TeV muon collider is a powerful machine for single Higgs precision which is comparable to or surpasses any Higgs factory proposed thus far in many channels~\cite{deBlas:2019rxi}.  Given the cleaner environment of a lepton collider, in many channels the precision achievable is not parametrically different than the original signal only estimates of~\cite{SmashersGuide}. Of course, there are some notable exceptions- $Z\gamma$ has large backgrounds which are difficult to remove, although the additional inclusion of the leptonic mode should be able to improve it somewhat. $t\bar{t}H$ is dominated by the much larger $t\bar{t}$ background that contributes due to our imperfect flavour tagging and jet clustering. The hadronic channels act as backgrounds to one another, especially impacting $gg$, $ZZ^*(4j)$ and $WW(4j)$.

There is still considerable room for improvement in all the studies presented here.  Optimized analysis techniques, improved detector designs, differential observables or even possible different $E_{CM}$ staging proposals all allow for potential improvements.  Ultimately, the goal is to be able to exploit the full potential of the improved energy and luminosity reach provided by a high energy muon collider.  As we have shown, two crucial components of detector design that can be focused on are forward muon tagging and increasing acceptance while maintaining BIB mitigation.  While BIB wasn't explicitly taken into account except for in Section~\ref{fullsim}, by using the default {\sc Delphes} design in combination with the angular distribution of the Higgs, it is clear that there is room for improving the acceptance of Higgs events with an improved detector or looser analysis cuts. This of course should not come at the expense of introducing additional BIB, but further studies optimizing the MDI for 10 TeV may naturally allow for this, in addition to incorporating more information from timing into particle reconstruction algorithms. 

Flavor tagging that has been done in this paper has not been optimized at all.  There is a rudimentary $b$-tagging algorithm based on the 3 TeV full simulation studies~\cite{Bartosik:2022ctn}, but in practice we have just used the tagging and fake rates from the {\sc Delphes} muon collider card augmented by ILC motivated working points for charm tagging at high energy.   We have also taken quite low working points for flavor tagging whereas a full analysis akin to that of CLIC \cite{Abramowicz:2016zbo} can better separate between $b\bar{b}$, $c\bar{c}$, and $gg$ final states.   It would be interesting to see how far a muon collider detector could be pushed for hadronic particle ID given the relatively clean environment.  While it may never reach the capabilities of a low energy $e^+e^-$ collider, which potentially can do strange tagging at a decent working point~\cite{Albert:2022mpk}, it is nevertheless interesting to find out what the boundary is for understanding the complementarity of various collider options.

Of course, there can also be difficulties encountered in more detailed studies.  As mentioned, we don't include the ``collinear'' backgrounds from $q/g$ components of the muon~\cite{Han:2021kes} when they are treated as part of a generalized collinear PDF of the muon.  There is a great deal of study still needed to understand the theoretical uncertainties in these predictions, as well as the ability to mitigate these backgrounds with e.g. forward tagging.  The formalism to simultaneously account for these contributions, contributions from low-virtuality $\gamma$'s, and detector mitigation techniques needs to be further developed to determine if the sensitivity is somewhat affected.

A clear avenue of research is to develop the understanding of how precise the Higgs width can be determined at a high energy muon collider.  For example techniques such as constraining the width through diboson measurements~\cite{Kauer:2012hd,Caola:2013yja} like at the LHC is possible and in progress.  While this isn't completely ``model-independent'', the cleaner environment of a muon collider can also allow for several observables to be combined which improves the measurement.  Of course, a high energy muon collider could also be preceded by an $e^+e^-$ collider or a muon collider at the Higgs pole, both of which would give strong constraints on the Higgs width from different techniques.

One particularly weak spot for high energy muon colliders is how to improve the measurement of $y_t$.  The LHC naturally provides stronger complementary measurements, but since the top quark plays a particularly interesting role in Higgs physics, it would be useful if there was a path to further improvement.  Combining other non-Higgs observables is a particularly interesting path, for example, in~\cite{SmashersGuide}, $W^+W^-\rightarrow t\bar{t}$ was used to constrain $y_t$ indirectly to a level similar to the HL-LHC projections.  However, this study was done without backgrounds, and it would be important to understand how far these types of observables could be improved at a high energy muon collider.   Another exciting option is to imagine different staging proposals to focus on top quark properties.  Muon collider staging thus far has focused on 125 GeV, 3 Tev, and 10+ TeV.  However, given the ability to go to higher energies, optimizing somewhere between 550 GeV to 1 TeV may be an interesting run plan before the ultimate 10+ TeV energy goal.  

While there is a great deal of work still to be performed to realize the dream of a high energy muon collider, the physics case for the Higgs is certainly strong.  This is especially true when coupled with multi-Higgs observables~\cite{Han:2020pif,Chiesa:2020awd} or a Higgs pole collider~\cite{deBlas:2022aow}, or imagining complementary datasets with the HL-LHC and $e^+e^-$ Higgs factories.  If the Higgs program of a high energy muon collider were the only physics goal, it would already provide strong motivation to pursue this direction.  However, it's also important to remember this is just one facet of a high energy muon collider program, and reaching the 10+ TeV scale allows for the greatest BSM physics reach of any proposed collider in numerous BSM physics scenarios~\cite{SmashersGuide}.  Most importantly, the proposed IMCC timeline~\cite{Adolphsen:2022bay} gives us the quickest path to this scale.

\acknowledgments

We would like to thank Dimitrios Athanasakos, Simone Pagan Griso, Samuel Homiller, Sergo Jindariani, Zhen Liu, Donatella Lucchesi, and  Lorenzo Sestini for useful conversations and details that enabled this study to be completed.  This work of PM and MF was supported in part by the National Science Foundation grant PHY-1915093.

\bibliographystyle{utphys}
\bibliography{bib}

\end{document}